\documentclass[aps,prd,floats,nofootinbib]{revtex4}
\usepackage{latexsym}
\usepackage{amssymb}
\usepackage[dvips]{graphicx}
\usepackage[applemac]{inputenc}
\usepackage[nottoc]{tocbibind} 
\usepackage{float}
\usepackage{fancyhdr}
\usepackage{amsfonts}
\usepackage{amsmath}
\usepackage{color}
\usepackage{mathrsfs}
\usepackage{bm}
\usepackage{epsfig}
\usepackage{feynmf}
\usepackage{multirow}
\def\beq{\begin{equation}}
\def\eeq{\end{equation}}
\def\beqa{\begin{eqnarray}}
\def\eeqa{\end{eqnarray}}
\newcommand{\lsim}{\mbox{\raisebox{-.6ex}{~$\stackrel{<}{\sim}$~}}}

\def\sss{\scriptscriptstyle}

\def\rmscr{\rm\scriptscriptstyle}
\def\newpara{\vskip 0.5cm\noindent}

\def\L{\left}
\def\R{\right}
\def\G{\,{\rm G}}
\def\cm{\,{\rm cm}}
\def\m{\,{\rm m}}
\def\s{\,{\rm s}}
\def\km{\,{\rm km}}
\def\kmps{\km\s^{-1}}

\def\yr{\,{\rm yr}}
\def\pc{\,{\rm pc}}
\def\kpc{\,{\rm kpc}}

\def\gev{\,{\rm GeV}}
\def\gevcc{\,{\rm GeV/cm^3}}

\def\pb{\,{\rm pb}}
\def\half{\frac{1}{2}}
\def\dhalf{\dfrac{1}{2}}
\def\exp{{\rm exp}}
\def\erf{{\rm erf}}

\def\threehalf{\frac{3}{2}}

\def\boldx{{\bf x}}
\def\boldv{{\bf v}}

\def\boldu{{\bf u}}

\def\mathcalR{{\mathcal R}}

\def\mathcalE{{\mathcal E}}

\def\bx{{\bf x}}

\def\bv{{\bf v}}
\def\bu{{\bf u}}

\def\min{{\rm min}}

\def\si{{\rm SI}}
\def\sd{{\rm SD}}

\def\sigmachin{\sigma_{\chi n}}
\def\sigmachip{\sigma_{\chi p}}

\def\sigmachipSI{\sigmachip^{\si}}
\def\sigmachinSI{\sigmachin^{\si}}

\def\sigmachipSD{\sigmachip^{\sd}}

\def\sigmachiA{\sigma_{\chi {\rmscr A}}}
\def\sigmachiASI{\sigma_{\chi {\rmscr A}}^{\si}}

\def\muchiA{\mu_{\chi {\rmscr A}}}

\def\muchip{\mu_{\chi {\rm p}}}

\def\muchii{\mu_{\chi i}}

\def\umin{u_{\rm min}}
\def\umax{u_{\rm max}}
\def\vmax{v_{\rm max}}

\def\vsun{\,v_\odot}
\def\wesc{w_{\rm esc}}
\def\wesccore{w_{\rm esc, core}}
\def\wescsurf{w_{\rm esc, surf}}

\def\vdisp{\langle v^2 \rangle^{1/2}}

\def\rhodm{\rho_{\rm\scriptscriptstyle DM}}
\def\rhovis{\rho_{\rm\scriptscriptstyle VM}}

\def\rhodmsun{\rho_{{\rm\scriptscriptstyle DM},\odot}}
\def\Phidm{\Phi_{\rm\scriptscriptstyle DM}}
\def\Phivis{\Phi_{\rm\scriptscriptstyle VM}}

\def\rsun{\,R_\odot}
\def\rt{r_t}

\def\mchi{m_\chi}
\def\mA{m_A}
\def\msun{\,M_\odot}
\def\dcdvdis{\dfrac{dC}{dV}}

\def\omegaminus{\Omega^-(w)}
\def\DelE{\Delta E}
\def\mup{\mu_+}
\def\mum{\mu_-}

\def\Csun{C_{\odot}}

\def\Gamsun{\Gamma_{\odot}}
\def\tsun{t_{\odot}}
\def\anu{\bar{\nu}}
\def\numu{\nu_{\mu}}
\def\numubar{\bar{\nu}_\mu}
\def\nutau{\nu_{\tau}}
\def\mtau{m_\tau}

\def\mb{m_b}
\def\mc{m_c}

\def\tautau{\tau^+\tau^-}
\def\bbarb{{\bar{b}}\,b}
\def\cbarc{{\bar{c}}\,c}
\def\qbarq{{\bar{q}}\,q}

\def\Emu{E_{\mu}}
\def\Ei{{E_i}}
\def\Eicore{E_i^{\rmscr core}}
\def\Ep{{E_+}}
\def\Em{{E_-}}

\def\Ec{E_c}
\def\Ed{E_d}

\def\dphiidEi{\frac{d\phi_i}{dE_i}}

\def\dNidEi{\frac{dN_i}{dE_i}}
\def\dNicoredEicore{\frac{dN_i^{\rmscr core}}{d\Eicore}}

\def\dNdEbyNhadron{\L(\dfrac{1}{N}\dfrac{dN}{d\Ed}\R)^{\hskip -0.05cm \rm hadron}}

\begin{document}
\title{\bf Neutrinos from WIMP annihilation in the Sun : 
Implications of 
a self-consistent model of the Milky Way's dark matter halo}
\author{Susmita Kundu\footnote{susmita.kundu@saha.ac.in} and  
Pijushpani 
Bhattacharjee\footnote{pijush.bhattacharjee@saha.ac.in}}
\affiliation{
AstroParticle Physics \& Cosmology Division and Centre for AstroParticle Physics,\\
Saha Institute of Nuclear Physics,~1/AF Bidhannagar,~Kolkata~700064.~India
}
\begin{abstract}
\noindent 
Upper limits on the spin-independent (SI) as well as spin-dependent (SD) elastic 
scattering cross sections of low mass ($\sim$ 2 -- 20 GeV) WIMPs (Weakly Interacting Massive 
Particles) with protons, imposed by the upper limit on the 
neutrino flux from WIMP annihilation in the Sun given by the Super-Kamiokande (S-K) 
experiment, and their compatibility with the ``DAMA-compatible" regions of the WIMP parameter 
space --- the regions of the WIMP mass versus cross section parameter space 
within which the annual modulation signal observed by the DAMA/LIBRA experiment is compatible 
with the null results of other direct detection experiments --- are  
studied within the frame work of a self-consistent model of the finite-size dark matter (DM) 
halo of the Galaxy. The halo model includes  the gravitational influence of 
the observed visible matter of the Galaxy on the phase space distribution function of the 
WIMPs constituting the Galaxy's DM halo in a self-consistent manner. Unlike in the  
``Standard Halo Model" (SHM) used in earlier analyses, the 
velocity distribution of the WIMPs in our model is non-Maxwellian, with a high-velocity cutoff 
determined self-consistently by the model itself. The parameters of the model are determined 
from a fit to the rotation curve data of the Galaxy. We find that, for our best fit halo 
model, for SI interaction, while the S-K upper limits do not place additional 
restrictions on the DAMA-compatible region of the WIMP parameter space if the WIMPs annihilate 
dominantly to $\bbarb$\ and/or $\cbarc$, portions of the DAMA-compatible region can be 
excluded if WIMP annihilations to $\tautau$ and $\nu\anu$ occur at larger than $35\%$ and 
$0.4\%$ levels, respectively. For SD interaction, on the other hand, 
the restrictions on the possible annihilation channels are much more 
stringent: they rule out the entire DAMA region if WIMPs annihilate to $\tautau$ and 
$\nu\anu$ final states at greater than $\sim\,$0.05\% and 0.0005\% levels, respectively, 
and/or to $\bbarb$ and $\cbarc$ at greater than $\sim\,$0.5\% levels. The very latest 
results from the S-K Collaboration [T.~Tanaka et al, Astrophys.~J.~{\bf 742}:78 (2011)] 
make the above constraints on the branching fractions of various WIMP annihilation channels 
even more stringent by roughly a factor of 3--4.     

\end{abstract}
\maketitle
\newpage
\section{Introduction}\label{sec:Intro}
\noindent
Weakly Interacting Massive Particles (WIMPs) (hereafter generically denoted by $\chi$) 
with masses $\mchi$
in the range of few GeV to few TeV are a natural candidate for the dark 
matter (DM) in the Universe; See e.g., 
Refs.~\cite{Jungman_etal_PhysRep_96,Bergstrom_00,Bertone_etal_PhysRep_05, 
Hooper_Profumo_KKDM_PhysRep_07,Bertone_ed_book_10} 
for reviews. Several experiments are currently engaged in efforts to 
directly detect such WIMPs by observing nuclear recoils due to 
scattering of WIMPs off nuclei in suitably chosen detector materials in 
underground laboratories. Recent results from some of these direct 
detection (DD) experiments, in particular the annual modulation of the 
nuclear recoil event rates reported by the DAMA/LIBRA 
collaboration~\cite{dama-libra} and the excess of low energy recoil  
events reported by the CoGeNT collaboration~\cite{cogent_2010}  have 
raised the interesting possibility~\cite{Hooper-Collar-Hall-McKinsey-Kelso_1007_1005,  
Fitzpatrick-Hooper-Zurek_1003_0014} that these events could be due to 
WIMPs of relatively low mass, approximately in the range $\sim$ 5--10 
GeV, interacting with nuclei with a WIMP-nucleon spin-independent 
elastic cross section in the region of few $\times10^{-4}\,$pb, without 
conflicting with the null results from other experiments such as 
XENON10~\cite{xenon10}, XENON100~\cite{xenon100-2010} and 
CDMS-II-Si~\cite{cdmsII-Si}. Earlier analyses (before the CoGeNT 
results~\cite{cogent_2010}) had 
also found similar compatibility of the DAMA/LIBRA annual modulation 
signal with the null results from other DD experiments; see, e.g., 
Refs.~\cite{Petriello-Zurek_08,Savage_etal_08,cbc_JCAP2010}.~\footnote{The question of 
compatibility of the DAMA/LIBRA and CoGeNT results with the null results of other experiments, 
however, remains controversial; see, e.g., the results of a recent reanalysis of the CDMS-II 
Germanium data with a lowered recoil-energy threshold of 2 
keV~\cite{cdmsII-Ge_2keV_1011_2482}, as well as the recent results from the 
XENON100 collaboration~\cite{xenon100-2011}, both of which claim to disfavor such a 
compatibility.}

\newpara
Scattering of WIMPs off nuclei can also lead to capture of 
the WIMPs by massive astrophysical bodies such as the Sun or the 
Earth if, after scattering off a nucleus inside the body, the velocity 
of the WIMP falls below the escape velocity of the body. The WIMPs so 
captured over the lifetime of the capturing body would gradually settle 
down to the core of the body where they would annihilate and produce 
standard model particles, e.g., $W^+W^-,\, Z^0Z^0,\, \tau^+\tau^-,\, 
t\bar{t},\, b\bar{b},\, c\bar{c},\,$ etc. Decays of 
these particles would then produce neutrinos, gamma rays, 
electrons-positrons, protons-antiprotons, etc.  
For astrophysical objects like the Sun or the Earth, only the neutrinos 
would be able to escape. Detection of these neutrinos by large neutrino 
detectors can, albeit indirectly, provide a signature of 
WIMPs. Although no detection has yet been reported, the Super-Kamiokande (S-K) 
detector, for example, has provided upper limits on the possible neutrino flux from 
WIMP annihilation in the Sun as a function of the WIMP 
mass~\cite{SuperK_limit_04,Wink_1104_0679,S-K_arxiv:1108.3384}. 
Similarly, the $\gamma$-rays produced in the annihilation of the WIMPs in 
suitable astrophysical environments with enhanced DM density but low 
optical depth to gamma rays, such as in the central region of our 
Galaxy, in dark matter dominated objects such as dwarf 
galaxies, and in clusters of galaxies,  
can offer a complimentary avenue of indirect detection (ID) of WIMPs.  
Although no unambiguous gamma ray signals of dark matter origin have been  
reported, a recent analysis~\cite{Hooper-Goodenough_1010_2752} of the spectral and 
morphological features of the gamma ray emission from the inner Galactic Center region 
(within a Galactocentric radius of $\sim175\pc$) measured by the Fermi Gamma-ray Space 
Telescope (FGST) seems to suggest the presence of a gamma ray emission component which 
is difficult to explain in terms of known sources and/or process 
of gamma ray production, but is consistent with that expected from annihilations of 
WIMPs of mass in the 7--9 GeV range (annihilating primarily to tau leptons) with a 
suitably chosen density and distribution of the dark matter in the Galactic Center region; 
see, however, Ref.~\cite{Boyarsky_etal_1012_5839} for a different view. 

\newpara
In this paper we focus on the neutrinos produced by annihilations of WIMPs in the core 
of the Sun, and study the constraints imposed on the WIMP mass  
vs.~WIMP-nucleon cross section, for low-mass ($\lsim20\gev$) WIMPs,  
from non-detection of such neutrinos. This is done within the context of a 
self-consistent model of the finite-size dark halo of the 
Galaxy~\cite{crbm_NewAstron_2007,cbc_JCAP2010} that includes the 
gravitational effect of the observed visible matter on the DM in a 
self-consistent manner, with the parameters of the model   
determined from fits to the rotation curve data of the 
Galaxy~\cite{Honma-Sofue_97,rc_60kpc_Xue_etal_08}. 

\newpara
The expected flux of neutrinos from the Sun due to WIMP annihilations 
depends on the rate at which WIMPs are captured by the Sun. The capture rate depends 
on the density as well as the velocity distribution of the WIMPs 
in the solar neighborhood as the Sun goes around the Galaxy. The density and velocity 
distribution of the WIMPs in the Galaxy are {\it a priori} 
unknown. Most earlier studies of neutrinos from WIMP capture and annihilation in 
the Sun have been done within the context of the so-called ``Standard Halo Model" 
(SHM) in which the DM halo of the Galaxy is described by a single component
isothermal sphere~\cite{Binney_Tremaine} with a Maxwellian velocity distribution of 
the DM particles in the Galactic rest 
frame~\cite{FFG_88,Jungman_etal_PhysRep_96,Lewin_Smith_96}). The velocity distribution is 
isotropic, and is usually truncated at a chosen value of the escape speed of the Galaxy. 
The density of DM in the solar neighborhood is typically taken to be in the range 
$\rhodmsun\sim0.3\pm 
0.1\gev/\cm^3$~\cite{Oort,Bahcall_84,Salucci_etal_1003_3101,Ling_etal_simulations_09}
\footnote{See, however, recent analyses~\cite{Catena-Ullio_0907.0018,McMillan_1102.4340} 
which claim a value closer to $0.4\gev/\cm^3$ with uncertainty $\lsim 10\%$.}.  
The velocity dispersion, $\vdisp$, the parameter characterizing the Maxwellian velocity 
distribution of the SHM,  
is typically taken to be $\sim270\kmps$. This follows from the relation~\cite{Binney_Tremaine},
${\vdisp}=\sqrt{\threehalf} v_{c,\infty}$, between the velocity
dispersion of the particles constituting a single-component
self-gravitating isothermal sphere and the
asymptotic value of the circular rotation speed, $v_{c,\infty}$,
of a test particle in the gravitational field of the isothermal sphere and assuming 
$v_{c,\infty}\approx v_{c,\odot}\approx 220\kmps$, where $v_{c,\odot}$
is the measured value of the circular rotation velocity of the Galaxy in
the solar neighborhood.~\footnote{A somewhat higher value of
$v_{c,\odot}\approx 250\kmps$, as suggested by a recent study~\cite{new_rot_speed}, would imply a
correspondingly higher value of ${\vdisp}_{\rm iso}\approx306\kmps$.} 
Neutrino flux from DM annihilation in the Sun for low mass WIMPs and the resulting constraints 
on WIMP properties from the Super-Kamiokande upper limits on such 
neutrinos have been studied within the context of the SHM in 
Refs.~\cite{Feng_0808_4151,Hooper_Petriello_Zurek_Kamion_PRD_09,Niro_0909_2348,Wink_1104_0679}, 
which showed that the Super-Kamiokande upper limits on the possible flux of neutrinos from 
the Sun place stringent restrictions on the DAMA region of the WIMP parameter space.  

\newpara
Whereas the SHM serves as a useful benchmark model, there are a number of reasons why 
the SHM does not provide a satisfactory description of the dynamics of the Galaxy. 
First, it does not take into account the modification of the phase space structure of 
the DM halo due to the significant gravitational effect of the observed visible matter on the 
DM particles inside and up to the solar circle. Second, the isothermal sphere model of the 
halo is infinite in extent and has a 
formally divergent mass, with mass inside a radius $r$, $M(r)\propto r$, as $r\to\infty$, and 
is thus unsuitable for representing a halo of finite size. Third, the procedure of 
truncating the Maxwellian speed distribution at a chosen value of the local (solar
neighborhood) escape speed is not a self-consistent one because the resulting 
speed distribution is not in general a self-consistent solution of the
steady-state collisionless Boltzmann equation describing a finite
system of collisionless DM particles. In addition, since the rotation curve for such a
truncated Maxwellian is, in general, not asymptotically flat, the
relation ${\vdisp}=\sqrt{\threehalf} v_{c,\infty}$ used to determine the
value of $\vdisp$ in the Maxwellian speed distribution of the
isothermal sphere, as done in the SHM, is not valid in general. Finally, recent numerical 
simulations~\cite{Ling_etal_simulations_09} seem to find that the velocity distribution of the 
Dark Matter particles deviates significantly from the usual Maxwellian form. These issues 
are further discussed in detail in Ref.~\cite{cbc_JCAP2010}, where we discussed a 
self-consistent model of the finite-size dark halo of the Galaxy that avoids the above 
mentioned inconsistencies of the SHM and also studied the 
constraints on WIMP properties from the results of the direct detection (DD) experiments 
within the context of this self-consistent halo model. It is of interest to extend this 
study to the case of indirect detection (ID) of WIMPs via neutrinos from WIMP annihilations 
in the Sun, which is the purpose of this paper.  

\newpara 
Our model of the phase space structure of the finite-size DM halo of the 
Galaxy is based on the so-called ``lowered" (or truncated) isothermal 
models (often called ``King models")~\cite{Binney_Tremaine} of the phase-space 
distribution function (DF) of collisionless particles. These models are proper self-consistent
solutions of the collisionless Boltzmann equation representing nearly
isothermal systems of finite physical size and mass. There are two important features of these 
models: First, at every location within the system a DM particle can have speeds up to a maximum
speed which is self-consistently determined by the model itself. A particle of maximum velocity 
at any location within the system can just reach its outer boundary, fixed by the 
truncation radius, a parameter of the model, where the DM density by construction vanishes.  
Second, the speed distribution of the particles constituting the system is non-Maxwellian. 
To include the gravitational effect of the observed visible matter on the DM particles, 
we modify the ``pure" King model DF 
by replacing the gravitational potential appearing in the King model DF by the total 
gravitational potential consisting of the sum of those due to DM and the observed visible 
matter. This interaction with the visible matter influences both the density profile and the
velocity distribution of the dark matter particles as compared to those for a ``pure" King 
model. In particular, the dark matter is pulled in by the visible matter, thereby  
increasing its central density significantly. 
When the visible matter density is set to zero and the truncation radius is set
to infinity, our halo model becomes identical to that of a single-component isothermal sphere 
used in the SHM. For further discussion of the model, see \cite{crbm_NewAstron_2007,cbc_JCAP2010}. 

\newpara
The DM distribution in the Galaxy may have significant amount of substructures which 
may have interesting effects on the WIMP capture and annihilation 
rates~\cite{Koushiappas-Kamion_PRL_09}. However, not much information, based 
on observational data, is available about the spatial 
distribution and internal structures of these substructures. As such, in this 
paper we shall be concerned only with the smooth component of the DM distribution 
in the Galaxy described by our self-consistent model mentioned above, the  
parameters of which are determined from the observed rotation curve data for the Galaxy. 

\newpara
The non-Maxwellian nature of the WIMP speed distribution in our halo model makes the 
calculation of the WIMP capture (and consequently annihilation) rate non-trivial since the 
standard analytical formula for the capture rate given by Gould~\cite{Gould} and Press and 
Spergel~\cite{Press-and-Spergel}, which is widely used in the literature, is not valid 
for the non-Maxwellian speed distribution in our halo model, and as such has to be 
calculated ab initio; see section \ref{sec:capture_annih_rate}. 

\newpara
We calculate the 90\% C.L.~upper limits on the WIMP-proton spin-independent (SI) as well as 
spin-dependent (SD) elastic cross sections as a function of the WIMP mass, for various WIMP 
annihilation channels, using the 90\% C.L. upper limits on the rates of upward-going 
muon events due to neutrinos from Sun derived from the results of 
S-K collaboration~(\cite{SuperK_limit_04}, \cite{Wink_1104_0679} and references 
therein).~\footnote{After the completion of the main calculations of the present work, new 
results of the S-K collaboration's search for upward-going muons due to neutrinos from 
Sun~\cite{S-K_arxiv:1108.3384} have appeared. We include, at the end of 
section \ref{sec:Results}, a discussion of the 
new results of Ref.~\cite{S-K_arxiv:1108.3384} and the resulting constraints on various 
WIMP annihilation channels.} We then study the consistency of those limits with the 
90\% C.L. ``DAMA-compatible" regions --- the regions of the 
WIMP mass versus cross section parameter space within which the annual modulation signal 
observed by the DAMA/LIBRA experiment~\cite{dama-libra} is compatible with the null results 
of other DD experiments --- determined within the context of our halo model~\cite{cbc_JCAP2010}.
We find that the requirement of such consistency imposes stringent restrictions on the 
branching fractions of the various WIMP annihilation channels. For example, in the case of 
spin-independent WIMP-proton interaction, while the S-K upper limits do not place additional 
restrictions on the DAMA-compatible region of the WIMP parameter space if the WIMPs annihilate 
dominantly to $\bbarb$,\ $\cbarc$, portions of the DAMA-compatible region can be excluded if 
WIMP annihilations to $\tautau$ and $\nu\anu$ occur at larger than 35\% and 0.4\% levels, 
respectively. In the case of spin-dependent interactions, on the other hand, the restrictions 
on the branching fractions of various annihilation channels are much more stringent. 
Specifically, they rule out the entire DAMA region if WIMPs annihilate to $\tautau$ and 
$\nu\anu$ final states at greater than $\sim\,$0.05\% and 0.0005\% levels, respectively, 
and/or to $\bbarb$ and $\cbarc$ at greater than $\sim\,$0.5\% levels.~\footnote{In the 
present paper, the CoGeNT results~\cite{cogent_2010} are not included in the analysis. 
Preliminary results of the analysis~\cite{cb_inprep_2011} to find the ``CoGeNT-compatible"  
region in the WIMP mass vs.~cross section plane {\it within the context of our 
halo model} indicates that its inclusion will not significantly change 
the above constraints on the branching fractions for the various annihilation 
channels.} The very latest results from the S-K Collaboration~\cite{S-K_arxiv:1108.3384}
make the above constraints on the branching fractions of various WIMP annihilation channels
even more stringent by roughly a factor of 3--4.

\newpara
The rest of the paper is organized as follows: In section \ref{sec:DF_model} we 
briefly describe the self-consistent model of the DM halo of the Galaxy. The formalism of 
calculating the WIMP 
capture and annihilation rates in the Sun within the context of our DM halo model,  
and that for calculating the resulting neutrino flux and event rate in the Super-Kamiokande 
detector, are discussed in sections \ref{sec:capture_annih_rate} and 
\ref{sec:neutrino_flux_event_rate}, respectively. Our results and the constraints 
on the WIMP properties implied by these results are described in  
section \ref{sec:Results}. The paper ends with a Summary in section \ref{sec:Summary}. 
\section{The self-consistent truncated isothermal model of the Milky Way's Dark Matter halo}
\label{sec:DF_model}
\noindent 
The phase space distribution function (DF) of the DM particles constituting a truncated isothermal  
halo of the Galaxy can be taken, in the rest frame of the Galaxy, to be of the ``King model" 
form~\cite{Binney_Tremaine,crbm_NewAstron_2007,cbc_JCAP2010}, 
\begin{equation}
f(\bx,\bv)\equiv f(\mathcalE)=\left\{ \begin{array}{ll}
\rho_1
(2\pi\sigma^2)^{-3/2}\left(e^{\mathcalE/\sigma^2} -1\right) &
\mbox{for $\mathcalE > 0$}\,,\\
0 & \mbox{for $\mathcalE\leq 0$}\,,
\end{array}
\right.
\label{eq:king_df}
\end{equation}
where
\begin{equation}
\mathcalE(\boldx)\equiv\Phi(\rt)-\left(\half v^2 +
\Phi(\boldx)\right)\equiv
\Psi(\boldx)-\half v^2\,,
\label{eq:relative_energy_def}
\end{equation}
is the so-called ``relative energy" and $\Psi(\boldx)=
-\Phi(\boldx)+\Phi(\rt)$ the ``relative potential", $\Phi(\boldx)$ being the total  
gravitational potential under which the particles move, with boundary condition $\Phi(0)=0$. 
The relative 
potential and relative energy, by construction, vanish at $|\boldx|=\rt$, the truncation 
radius, which represents the outer edge of the system where the particle density vanishes. 
At any location $\boldx$ the maximum speed a particle of the system can
have is
\beq
\vmax (\boldx)=\sqrt{2\Psi(\boldx)}\,,
\label{eq:vmax}
\eeq
at which the relative energy $\mathcalE$ and, as a consequence, the DF
(\ref{eq:king_df}), vanish. The model has three parameters, namely, 
$\rho_1$, $\sigma$ and $\rt$. Note that the parameter $\sigma$ in the King model is not 
same as the usual velocity dispersion parameter of the isothermal phase space 
DF~\cite{Binney_Tremaine}. Also, in our calculations below, we shall use the parameter 
$\rhodmsun$, the value of the DM density at the location of the Sun, in place of the 
parameter $\rho_1$.     

\newpara
Integration of $f(\bx,\bv)$ over all velocities gives the DM density at the 
position $\boldx$: 
\beqa
\rhodm(\boldx) & = &
\frac{\rho_1}{\left(2\pi\sigma^2\right)^{3/2}}
\int_0^{\sqrt{2\Psi(\boldx)}} dv\, 4\pi v^2
\left[\exp\left(\frac{\Psi(\boldx)-v^2/2}{\sigma^2}\right) -
1\right]  \\
 & = &
\rho_1\left[\exp\left(\frac{\Psi(\boldx)}{\sigma^2}\right)\,
\erf\left(\frac{\surd{\Psi(\boldx)}}{\sigma}\right) -
\sqrt{\frac{4\Psi(\boldx)}{\pi\sigma^2}}\left(1 +
\frac{2\Psi(\boldx)}{3\sigma^2}\right)\right]\,,
\label{eq:rho_dm}
\eeqa
which satisfies the Poisson equation
\beq 
\nabla^2 \Phidm(\boldx) = 4\pi G \rhodm(\boldx)\,,
\label{eq:poisson_eqn_DM}
\eeq
where $\Phidm$ is the contribution of the DM component to the total gravitational potential, 
\beq
\Phi(\boldx) = \Phidm(\boldx) + \Phivis(\boldx)\,,
\label{eq:Phitotal}
\eeq
{\it in presence of the visible matter} (VM) whose gravitational potential, $\Phivis$, 
satisfies its own Poisson equation, namely, 
\beq
\nabla^2\Phivis(\boldx)=4\pi G \rhovis(\boldx)\,. 
\label{eq:poisson_eqn_VM}
\eeq
We choose the boundary conditions
\beq
\Phidm(0)=\Phivis(0)=0\,,\,\,\,\,\, {\rm and} \,\,\,\,
\left(\nabla\Phidm\right)_{|\boldx|=0} =
\left(\nabla\Phivis\right)_{|\boldx|=0} = 0\,.
\label{eq:boundary_cond}
\eeq
The mass of the system, defined as the total mass contained within $\rt$, is given by 
$\G M(\rt)/\rt = \left[\Phi(\infty)-\Phi(\rt)\right]$. Note
that, because of the chosen boundary condition $\Phi(0)=0$, 
$\Phi(\infty)$ is a non-zero positive constant. 

\newpara
Since the visible matter distribution $\rhovis(\boldx)$, and hence the
potential $\Phivis(\boldx)$, are known from observations and modeling, 
the solutions of equation (\ref{eq:poisson_eqn_DM}) together with 
equations (\ref{eq:rho_dm}), (\ref{eq:Phitotal}) and the boundary conditions 
(\ref{eq:boundary_cond}),
give us a three-parameter family of self-consistent pairs of 
$\rhodm(\boldx)$ and $\Phidm(\boldx)$ for chosen values of the
parameters ($\rho_1\,, \sigma\,, r_t$). The values of these parameters for the Galaxy 
can be determined by comparing the theoretically calculated rotation curve, $v_c(R)$, given by  
\begin{equation}
v_c^2 (R) = R\frac{\partial}{\partial R}\Big[\Phi(R,z=0)\Big] =
R\frac{\partial}{\partial R}\Big[\Phidm(R,z=0)+\Phivis(R,z=0)\Big]\,,
\label{eq:v_c_def}
\end{equation}
with the observed rotation curve data of the Galaxy. (Here  
$R$ is Galactocentric distance on the equatorial plane and $z$ is 
the distance normal to the equatorial plane.) This procedure was described in detail in 
Refs.~\cite{crbm_NewAstron_2007,cbc_JCAP2010} where, for the visible matter density 
distribution described there, 
we determined the values of the parameters $\rt$ and $\sigma$ that 
gave reasonably good fit to the rotation curve data of the
Galaxy~\cite{Honma-Sofue_97,rc_60kpc_Xue_etal_08} for each of the three chosen values 
of the parameter $\rhodmsun =$ 0.2, 0.3 and 0.4 $\gev/\cm^3$. These models are summarized in 
Table \ref{Table:good_fit_models}, which we use for our calculations in this paper.  
%
\begin{table}[h]
\centering
\begin{tabular}{|c|c|c|c|c|c|}
\hline
Model & $\rhodmsun$ & $\rt$ & $\sigma$ \\
 & $(\gevcc)$ & $(\kpc)$ & $(\kmps)$\\
\hline
\hline
 M1 & $~0.2~$ & $~120.0~$ & $~300.0~$ \\
\hline
 M2 & $~0.3~$ & $~80.0~$ & $~400.0~$ \\
\hline
 M3 & $~0.4~$ & $~80.0~$ & $~300.0~$ \\
\hline
\end{tabular}
\caption{Parameters of our self-consistent model of the Milky Way's Dark Matter halo that 
give good fits to the Galaxy's rotation curve data, for the three chosen values of the DM 
density at the solar neighborhood}
\label{Table:good_fit_models}
\end{table}
The density profiles, mass profiles, velocity distributions of the DM particles and the resulting 
rotation curves in each of these models are discussed in detail in Ref.~\cite{cbc_JCAP2010}. 

\newpara
With our halo model specified, we now briefly review the basic formalism of calculating the 
WIMP capture and annihilation rates within the context of our halo model.  
\section{Capture and Annihilation Rates}\label{sec:capture_annih_rate}
\noindent
The capture rate per unit volume at radius $r$ inside the Sun can be written 
as ~\cite{Gould,Press-and-Spergel}
\begin{equation}\label{CapRate_dCdV}
\dcdvdis (r) =\int d^3\bu \,\frac{\tilde{f}(\bu)}{u}\,w\omegaminus\,,
\end{equation}
where $\tilde{f}(\bu)$ is the WIMP velocity distribution, as measured in the Sun's rest frame, 
in the neighborhood of the Sun's location in the Galaxy,  
and $w(r)=\sqrt{u^2+\wesc(r)}$ is the WIMP's speed at the radius $r$ inside the Sun, 
$\wesc(r)$ being the escape speed at that radius inside the Sun, which is related to the 
escape speed at the Sun's core, $\wesccore\approx 1354\kmps$, and that at its surface, 
$\wescsurf\approx 795\kmps$, by the approximate relation 
\begin{equation}
{\wesc}^2(r)=({\wesccore})^2-\dfrac{M(r)}{\msun}\L[{(\wesccore})^2-{(\wescsurf})^2 \R]\,.
\label{sun_escape_speed_wesc}
\end{equation}
The quantity   
$\omegaminus$ is the capture probability per unit time, which is just the product 
of the scattering rate and the conditional probability that after a scattering 
the WIMP's speed falls below the escape speed. 

\newpara
We shall here consider only the elastic scattering of the WIMPs off nuclei. The dominant 
contribution to the WIMP capture rate will come from the WIMPs scattering off 
hydrogen and helium nuclei. While for hydrogen, both spin-independent (SI) 
as well as spin-dependent (SD) cross sections, $\sigmachipSI$ and $\sigmachipSD$, respectively, 
will contribute, only SI cross section for helium is relevant. (We neglect here the small 
contribution from $^3{\rm He}$). In general, the effective momentum-transfer ($q$) dependent 
WIMP-nucleus SI scattering cross section, $\sigmachiASI(q)$, can be written in the usual way  
in terms of the ``zero-momentum" WIMP-proton (or WIMP-neutron) effective cross section, 
$\sigmachipSI=\sigmachinSI$, as 
\begin{equation}
\sigmachiASI (q)=\dfrac{\muchiA^2}{\muchip^2}\sigmachipSI~A^2 \L|F(q^2)\R|^2\,,
\label{eq:sigmachiASI_q}
\end{equation}
where $A$ is the number of neutrons plus protons in the nucleus, 
$\muchiA$ and $\muchip$ are the reduced masses of WIMP-nucleus and WIMP-proton systems, 
respectively, with $\muchii=(m_i~ \mchi)/(m_i+\mchi)$, and $F(q^2)$ is 
the nuclear form-factor (with $F(0)=0$) which can be chosen to be of the 
form~\cite{Jungman_etal_PhysRep_96} 
\begin{equation}\label{form_factor}
\L|F(q^2)\R|^2=\exp\L(-\dfrac{q^2R^2}{3\hbar^2}\R)=\exp\L(-\dfrac{\DelE}{E_0}\R)\,. 
\end{equation}
Here $R\sim \L[0.91\L(\dfrac{\mA}{\gev}\R)^{1/3} + 0.3\R] \times 10^{-13}\cm$ is the 
nuclear radius and $E_0\equiv 3\hbar^2/(2\mA R^2)$ is the characteristic 
nuclear coherence energy, $\mA$ being the mass of the nucleus.   

\newpara
With the above form of the nuclear form factor, the kinematics of the capture process~\cite{Gould} 
allows us to write the capture probability per unit time, $\omegaminus$, as 
\begin{eqnarray} 
\omegaminus&=&\frac{n_A~\sigmachiA}{w}~\frac{2E_0}{\mchi}~\frac{\mup^2}{\mu}
\L[\exp\L(-\frac{\mchi u^2}{2E_0}\R)-\exp\L(-\frac{\mchi w^2}{2E_0}~\frac{\mu}{\mup^2}\R)\R]
~\theta\L(\frac{\mu}{\mup^2}-\frac{u^2}{w^2}\R)\,,
\label{capture_prob_general}
\end{eqnarray}
where $n_A$ is the number density of the scattering nuclei at the radius $r$ inside the Sun, and 
$\mu \equiv \dfrac{\mchi}{\mA}$ , $\mu_{\pm}\equiv \dfrac{\mu \pm 1}{2}$. 
The $\theta$ function ensures that those particles which do not lose sufficient amount of 
energy to be captured are excluded.

\newpara
We shall use Equation (\ref{capture_prob_general}) to calculate $\omegaminus$ for helium ($A=4$). 
For hydrogen, however, there is no form-factor suppression, and the expression for $\omegaminus$ is 
simpler: 
\beq
{\rm Hydrogen: } \,\,\,\,\,\,\,\,\,\,\,\, 
\omegaminus = \frac{\sigmachip n_H}{w}\L(\wesc^2-\frac{\mum^2}{\mu} u^2\R)
                 \theta\L(\wesc^2-\frac{\mum^2}{\mu}u^2\R)\,,
\label{capture_prob_hydrogen}
\eeq
where $n_H$ is the density of hydrogen (proton) at the radius $r$ inside the Sun. 
Note that in equations (\ref{capture_prob_general}) and (\ref{capture_prob_hydrogen}), 
the quantities $w$, $\wesc$, $n_A$ and $n_H$ are functions of $r$.   

\newpara
The WIMP velocity distribution appearing in equation~(\ref{CapRate_dCdV}) is related to 
the phase space DF defined in equation (\ref{eq:king_df}) (valid in the rest frame of Galaxy) 
by the Galilean transformation  
\beq
\tilde{f}(\bu) = \frac{1}{\mchi}f\left(\boldx=\boldx_\odot,\boldv=\boldu+\boldv_\odot\right)\,,
\label{eq:ftilde_f_relation}
\eeq
where $\boldx_\odot$ represents the sun's position in the Galaxy ($R=8.5\kpc, z=0$)
and $\boldv_\odot$ is the Sun's velocity vector {\it in
the Galaxy's rest frame}. Note that Gould's original calculations and the final formula 
for the WIMP capture rate given in Ref.~\cite{Gould}, which are widely used in the literature, 
use a Maxwellian velocity distribution of the WIMPs in the Galaxy, and as such, cannot 
directly be used here since the WIMP velocity distribution in our case is non-Maxwellian. 
In particular, note that the DF $f$ of equation (\ref{eq:king_df}) vanishes for
speeds $v\ge\vmax$ defined in equation (\ref{eq:vmax}). Consequently,  
equation (\ref{CapRate_dCdV}) above can be written as  
\begin{equation}
\dcdvdis (r) =\frac{2\pi}{\mchi}\int_{-1}^{1} d(\cos\theta) \int_{\umin=\vsun}^{\umax(\cos\theta)}
u~du f\left(\boldx=\boldx_\odot,\boldv=\boldu+\boldv_\odot\right)\,
w\omegaminus\,,
\label{CapRate_dCdV_2}
\end{equation}
where $\vsun\approx 220$ -- $250\kmps$ is the Sun's circular speed in the Galaxy, and $\umax$ is 
given by the positive root of the quadratic equation 
\beq
\umax^2 + \vsun^2 + 2\umax\vsun\cos\theta = 2\Psi(\boldx=\boldx_\odot)\,.
\eeq
The total WIMP capture rate by the Sun, $C_\odot$, is given by 
\begin{equation}
C_\odot=\int_0^{\rsun}\, 4\pi r^2 dr\,\dfrac{dC(r)}{dV}\,,
\label{CapRate_total}
\end{equation}
where $\rsun$ is the radius of the Sun. 

\newpara
In this work we shall neglect the effect of evaporation of the captured WIMPs from the 
Sun~\footnote{Note, however, that evaporation may not be negligible   
for WIMP masses below $\sim 4\gev$ depending on the magnitude of the 
annihilation cross section~\cite{Hooper_Petriello_Zurek_Kamion_PRD_09}.}, 
and make the standard assumption that the capture and annihilation processes have reached an 
approximate equilibrium state over the long lifetime of the solar system 
($\tsun \sim 4.2$ billion yrs). Under these assumptions, the total annihilation rate of WIMPs in 
the Sun is simply related to the total capture rate by the relation  
\begin{equation}
\Gamsun\approx \half\Csun
\label{AnnRate}
\end{equation}
\section{Neutrino flux from WIMP annihilation in the Sun and event rate in the detector}
\label{sec:neutrino_flux_event_rate}
\subsection{The neutrino energy spectrum}
\label{subsec:neutrino_energy_spect}
\noindent 
In this subsection we collect together the known results for the energy spectra of neutrinos 
emerging from the Sun, for various WIMP annihilation channels 
\cite{Jungman-Kamion_PRD_1995,Cirelli_etal_05_08,Gaisser-Steigman-Tilav_86,Ritz-Seckel_88}, 
for use in the calculations described in the next subsection. 

\newpara
The differential flux of muon neutrinos  observed at Earth is~\cite{Jungman-Kamion_PRD_1995} 
\begin{equation}
\L(\dphiidEi\R)=\dfrac{\Gamsun}{4 \pi D^2}\sum_F B_F \L(\dNidEi\R)_F\,, \hskip 1cm 
(i=\numu, \numubar) 
\label{nu_flux_at_earth}
\end{equation}
where $\Gamsun$ is the rate of WIMP annihilation in the Sun, $D$ is the Earth-Sun distance, 
$F$ stands for the possible annihilation channels, $B_F$ is the branching ratio for the 
annihilation channel $F$ and $\L(\dNidEi\R)_F$ is the differential energy spectrum of the neutrinos 
of type $i$ emerging from the Sun resulting from the particles of annihilation channel $F$ injected 
at the core of the Sun. WIMPs can annihilate to all possible standard model particles e.g. 
$e^+e^-,\,\mu^+\mu^-,\,\tautau,\,\nu_e\anu_e,\,\numu\anu_\mu,\,\nutau\anu_\tau,\,  
\qbarq$ pairs and also gauge and higgs boson pairs ($W^+W^-,\, Z\bar{Z},\, h\bar{h}$), etc. 
In this paper we are only interested in low mass ($\sim 2$ - $20\gev$) WIMPs. Therefore,  
we will not consider WIMP annihilations to higgs and gauge boson pairs and top quark pairs. 
Light quarks like $u$, $d$, $s$ contribute very little  
to the energetic neutrino flux~\cite{Cirelli_etal_05_08}, and are not considered. 
The same is true for muons. So, in this paper we consider only the channels  
$\tautau\,, \,\bbarb\,, \,\cbarc\,\,$ and $\anu\nu$. 

\newpara
The neutrino energy spectra, $\L(\dNidEi\R)_F\,$, have been calculated 
numerically (see, e.g., ~\cite{Cirelli_etal_05_08,WimpSim}) by considering all the 
details of hadronization of quarks, energy loss of the resulting heavy hadrons, neutrino 
oscillation effects, neutrino energy loss due to neutral current interactions and absorption 
due to charged current interactions with the solar medium, $\nutau$ regeneration, etc.  
However, the numerical results in \cite{Cirelli_etal_05_08,WimpSim} are given for WIMP 
masses $\mchi \ge 10\gev$, and it is not obvious if those are valid for lower WIMP masses 
which are of our primary interest in this paper. In any case, given the presence of other 
uncertainties in the problem, particularly those associated with astrophysical quantities 
such as the local density of dark matter and its velocity distribution, we argue that it is 
good enough to use --- as we do in this paper --- approximate analytical expressions for the 
neutrino spectra available in the 
literature~\cite{Jungman-Kamion_PRD_1995,Gaisser-Steigman-Tilav_86,Ritz-Seckel_88}. 
We are interested in the fluxes of muon neutrinos and antineutrinos, for which we use the 
analytic expressions given in Ref.~\cite{Jungman-Kamion_PRD_1995}, which neglect neutrino 
oscillation effects. By comparing with the neutrino fluxes obtained from detailed numerical 
calculations~\cite{Cirelli_etal_05_08}, we find that for small WIMP masses below $\sim20\gev$ 
(the masses of our interest in this paper), the analytic expressions for the muon neutrino 
fluxes given in \cite{Jungman-Kamion_PRD_1995} match with the results of detailed numerical 
calculations~\cite{Cirelli_etal_05_08} to within a few percent.   
 
\newpara 
The main effect of the 
interaction of the neutrinos with the solar medium is that~\cite{Ritz-Seckel_88} 
a neutrino of type $i\, (=\numu, \numubar$) injected at the solar core with energy 
$\Eicore$ emerges from the Sun with an energy $\Ei$ given by  
\beq
\Eicore = E_i/(1-E_i\, \tau_i)\,,
\label{Ei-Eicore relation}
\eeq 
and with probability 
\beq
P_i=(1+\Eicore\tau_i)^{-\alpha_i} = (1-\Ei\tau_i)^\alpha\,,
\label{nu_core_surface_prob}
\eeq
with 
\beq
\alpha_{\numu}=5.1\,, \,\,\alpha_{\numubar}= 9.0\,, \,\,\, 
\tau_{\numu}=1.01 \times 10^{-3}\gev^{-1}\,, \,\, {\rm and}\,\,\,\, 
\tau_{\numubar}=3.8 \times 10^{-4}\gev^{-1}\,. 
\label{nu flux: alpha and tau values}
\eeq
Below we write down the expressions for the energy spectra of neutrinos emerging from the Sun 
for the four annihilation channels considered in this paper: 
\subsubsection[tau decay]{\bf $\tautau$ channel~:~ Neutrinos from decay of $\tau$ leptons 
$(\tau \rightarrow \mu \numu \nutau)$}\label{subsubsec:tau decay}
\noindent 
For this channel, the spectrum of muon-type neutrinos at the solar surface,  
including the propagation effects 
in the solar medium, can be written as~\cite{Jungman-Kamion_PRD_1995}
\beq
\L(\dNidEi\R)_{\tautau} =\, \L(1-E_i \tau_i\R)^{(\alpha_i-2)}
\L(\dNicoredEicore\R)_{\tautau} 
\,, \hskip 1cm (i=\numu\,, \numubar)
\label{numu flux:tau decay:solar surface} 
\eeq
where the relationship between $\Ei$ and $\Eicore$, and the values of $\alpha_i$ and $\tau_i$,  
are as given by equations (\ref{Ei-Eicore relation}) and (\ref{nu flux: alpha and tau values}), 
respectively, and 
\beq
\L(\dNicoredEicore\R)_{\tautau}= 
\dfrac{48\,\Gamma_{\tau  \rightarrow  \mu \numu \nutau}}{\beta \gamma \mtau^4}~\L(\dhalf 
\mtau \L({\Eicore}\R)^2-\dfrac{2}{3}\L({\Eicore}\R)^3\R)_{\Em}^{{\min}\L(\half\mtau,\Ep\R)}\,
\label{numu flux:tau decay:solar core}
\eeq
is the neutrino spectrum due to decay of the $\tau$-leptons injected at the solar core by WIMP 
annihilations. Here  
$\Gamma_{\tau  \rightarrow \mu \numu \nutau}=0.18\,,\,$ and 
$E_{\pm}=\dfrac{\Eicore}{\gamma\L(1\mp\beta\R)}\,$ with 
$\,\gamma = \L(1-\beta^2\R)^{-1/2}=\,\mchi/\mtau\,,$ $\mtau$ being the $\tau$-lepton mass.  

\newpara
Note that the $\nutau$s produced from $\tau$ decay may again produce $\tau$s by charged 
current interactions in the solar medium, and these secondary $\tau$s can decay to give 
secondary $\numu$s. But these $\numu$s would be of much lower energy compared to the 
primary $\numu$s from $\tau$ decay, and are not considered here. 
\subsubsection[b decay]{\bf $\bbarb$ channel~:~ Neutrinos from decay of $b$-quark hadrons 
$(b\rightarrow c \mu\numu)$}\label{subsubsec:b decay}
\noindent
The treatment is similar to the case of $\tau$ decay described above. However, here 
the hadronization of quarks and stopping of heavy hadrons in the solar medium 
have to be taken into account. The resulting spectrum of muon-neutrinos emerging from the Sun 
is given by~\cite{Jungman-Kamion_PRD_1995}
\beq
\L(\dNidEi\R)_{\bbarb}=\, 
\int_{\mb}^{E_0}\dNdEbyNhadron\hskip -.7cm (E_0,\Ed)
\L(1-\Ei \tau_i\R)^{(\alpha_i-2)}\L(\dNicoredEicore\R)_{\bbarb}\hskip -0.3cm 
\L(\Ed, \Eicore \R) d\Ed\,, \hskip 0.5cm (i=\numu\,, \numubar)
\label{numu flux:b decay:solar surface} 
\eeq
where $\mb$ is the $b$-quark mass, $E_0\approx 0.71\mchi$ is the initial energy 
of the b-quark hadron (the fragmentation function is assumed to be a sharply peaked 
function~\cite{Jungman-Kamion_PRD_1995}),      
\beq
\L(\dNicoredEicore\R)_{\bbarb}\hskip -0.3cm \L(\Ed, \Eicore \R) = 
\dfrac{48\,\Gamma_{b  \rightarrow  \mu \numu X}}{\beta \gamma \mb^4}~\L(\dhalf 
\mb\L({\Eicore}\R)^2-\dfrac{2}{3} \L({\Eicore}\R)^3\R)_{\Em}^{{\min}\L(\half\mb,\Ep\R)}
\label{numu flux:b decay:solar core}
\eeq
is the neutrino spectrum resulting from decay of the $b$-quark hadron injected at the 
solar core, and 
\beq
\dNdEbyNhadron\hskip -.7cm \L(E_0,\Ed\R) = \dfrac{\Ec}{\Ed^2}~\exp\L[\dfrac{\Ec}{E_0}-
\dfrac{\Ec}{\Ed}\R]\,,
\label{numu flux: hadron's decay energy distribution}
\eeq
with $\Ec\approx 470\gev$, is the distribution of the hadron's energy at the time of 
its decay if it is produced with an 
initial energy $E_0$. In equation (\ref{numu flux:b decay:solar core}),  
$\Gamma_{b  \rightarrow \mu \numu X}=0.103\,$ is the branching ratio for 
inclusive semi-leptonic decay of $b$-quark hadrons to muons~\cite{PDG}, and 
$E_{\pm}=\dfrac{\Eicore}{\gamma\L(1\mp\beta\R)}\,$ with 
$\,\gamma = \L(1-\beta^2\R)^{-1/2}=\,\Ed/\mb\,$. 
%
\subsubsection[c decay]{\bf $\cbarc$ channel~:~ Neutrinos from decay of $c$-quark hadrons 
$(c\rightarrow s \mu\numu)$}\label{subsubsec:c decay}
\noindent
Again, this is similar to the case of $b$-decay discussed above, except that the kinematics of 
the process is slightly different. The resulting muon 
neutrino spectrum is given by~\cite{Jungman-Kamion_PRD_1995} 
\beq
\L(\dNidEi\R)_{\cbarc} =\, 
\int_{\mc}^{E_0}\dNdEbyNhadron\hskip -.7cm (E_0,\Ed)
\L(1-\Ei \tau_i\R)^{(\alpha_i-2)}\L(\dNicoredEicore\R)_{\cbarc}\hskip -0.3cm 
\L(\Ed, \Eicore \R) d\Ed\,, \hskip 0.5cm (i=\numu\,, \numubar)
\label{numu flux:c decay:solar surface} 
\eeq
where $\mc$ is the $c$-quark mass, $E_0\approx 0.55\mchi$ is the initial energy 
of the charmed hadron,
\beq
\L(\dNicoredEicore\R)_{\cbarc}\hskip -0.3cm \L(\Ed, \Eicore \R) = 
\dfrac{8\,\Gamma_{c  \rightarrow  \mu \numu X}}{\beta \gamma \mc^4}~\L(\threehalf 
\mc \L({\Eicore}\R)^2-\dfrac{4}{3} \L({\Eicore}\R)^3\R)_{\Em}^{{\min}\L(\half\mc,\Ep\R)}
\label{numu flux:c decay:solar core}
\eeq
is the neutrino spectrum resulting from decay of the $c$-quarks injected at the 
solar core, with 
$\Gamma_{c  \rightarrow \mu \numu X}=0.13\,$, \ \ 
$E_{\pm}=\dfrac{\Eicore}{\gamma\L(1\mp\beta\R)}\,$, \  
$\,\gamma = \L(1-\beta^2\R)^{-1/2}=\,\Ed/\mc\,$,  
and $\dNdEbyNhadron\hskip -.7cm \L(E_0,\Ed\R)$ is given by equation 
(\ref{numu flux: hadron's decay energy distribution}) with $\Ec\approx 250\gev$ for 
$c$-quark. 
\subsubsection[prompt nu]{\bf $\nu \anu$ channel~:~ $(\chi\chi \to \numu \numubar)$}
\label{subsubsec:prompt nu}
\noindent
In this case the spectrum of muon neutrinos emerging from the Sun is simply given by 
\beq
\L(\dNidEi\R)_{\nu \anu} =\, \L(1-E_i \tau_i\R)^{(\alpha_i-2)}
\L(\dNicoredEicore\R)_{\nu \anu} 
\,, \hskip 1cm (i=\numu\,, \numubar)
\label{numu flux:prompt nu:solar surface} 
\eeq
where 
\beq
\L(\dNicoredEicore\R)_{\nu \anu} = 
\delta\L(\Eicore - \mchi\R) \equiv \L(1+\mchi\tau_i\R)^{-2}
\delta\L(\Ei - \frac{\mchi}{1+\mchi\tau_i}\R)\,.
\label{numu flux:prompt nu:solar core} 
\eeq
\subsection{Calculation of Event Rates in the Super-Kamiokande detector}
\label{subsec:EventRate}
\noindent
The rate of neutrino induced upward-going muon events, $\mathcalR$, in the S-K detector due to 
$\numu$s and $\numubar$s from WIMP annihilation in the Sun can be written 
as 
\beq
\mathcalR = \dhalf \sum_{i=\numu\,, \numubar}
\int \int\,\dphiidEi\,\dfrac{d\sigma_{i{\sss N}}}{dy}(\Ei,y)\, V_{\rm eff} (\Emu)  
n_p^{\hskip-0.07cm\rmscr water} d\Ei \, dy\,,
\label{event_rate_S-K}
\eeq
where $\dphiidEi$ is the differential flux of the neutrinos given by equation  
(\ref{nu_flux_at_earth}), $\dfrac{d\sigma_{i{\sss N}}}{dy}$ are the relevant neutrino-nucleon 
charged current differential cross sections, $(1-y) \L(=E_\mu/E_i\R)$ is the fraction of the 
neutrino energy transferred to the the muon, $V_{\rm eff} (\Emu)$ is the effective volume of 
the detector, and $n_p^{\hskip-0.07cm\rmscr water}$ is the number density of protons in water 
(= Avogadro number). The S-K Collaboration imposed a cut on the   
upward-going muon path-length of $>$ 7 meters in the inner detector 
which has an effective area of $A_{\rm eff}\approx 900 \m^2$ and height $\approx 36.2\, \m$.  
This 7-meter cut on the muon track length can be effectively taken into account by 
setting  
$V_{\rm eff}=0$ if the effective water-equivalent muon range, $R_\mu (\Emu)\approx 5 \, 
{\rm meters}\,\times (\Emu/\gev)\,$, is less than 7 meters, and 
$V_{\rm eff}\,= A_{\rm eff}\, \times \L[R_\mu (\Emu)\,+\,\L(36.2-7\R)\,{\rm meters}\R]$ 
otherwise~\cite{Hooper_Petriello_Zurek_Kamion_PRD_09}. The factor of $1/2$ accounts for the 
fact that only up-going muon events were considered in order to avoid the background due to 
down-going muons produced due to cosmic ray interactions in the Earth's atmosphere. 

\newpara
The S-K muon events were broadly classified into three categories~\cite{SuperK_Ashie}, 
namely, (i) Fully Contained~(FC), (ii) Stopping~(S) and (iii) Through-Going~(TG) events. For 
$\numu$ energy $\lesssim 4 \gev$ the events are predominantly of FC type, whereas for 
$\numu$ energy $\gtrsim 8 \gev$ the events are predominantly of TG type. Assuming that 
annihilation of the WIMP of mass $\mchi$ produces neutrinos of typical energy  $\sim 
(\frac{1}{3} - \frac{1}{2})\mchi$, we can  roughly divide the $\mchi$ range of our interest 
into three regions according to the resulting muon event types namely, (i) $2\lesssim \mchi 
\lesssim 8 \gev $ (FC), (ii)~$8\lesssim \mchi \lesssim 15 \gev $ (FC + S) and   (iii) 
$15\gev\lesssim \mchi$ (FC + S + TG).

\newpara
To set upper limits on the WIMP elastic scattering cross section as a function of WIMP mass 
for a given annihilation channel, we use the following 90\% C.L.  upper limits 
~\cite{SuperK_limit_04,Wink_1104_0679} on the rates of the upgoing muon events of the three 
different types mentioned above:  
\beqa \nonumber
\mathcalR^{90\% {\rm C.L.}}_{\rm FC} & \simeq & 13.8 ~\yr^{-1}\,,\\ \nonumber
\mathcalR^{90\% {\rm C.L.}}_{\rm S} & \simeq & 1.24 ~\yr^{-1}\,,\\ 
\label{fc_s_tg_upper_limits}
\mathcalR^{90\% {\rm C.L.}}_{\rm TG} & \simeq & 0.93 ~\yr^{-1}\,.
\eeqa
The upper limits on the WIMP-nucleon elastic scattering cross section so derived  are then 
translated into upper limits on the branching fractions of various annihilation channels by 
demanding the consistency of DAMA-compatible region of the WIMP parameter space with S-K 
upper limits. These limits are discussed in the next section. 
  
\section{Results and Discussions}\label{sec:Results}
\begin{figure}[h]
\centering
\begin{tabular}{cc}
\epsfig{file=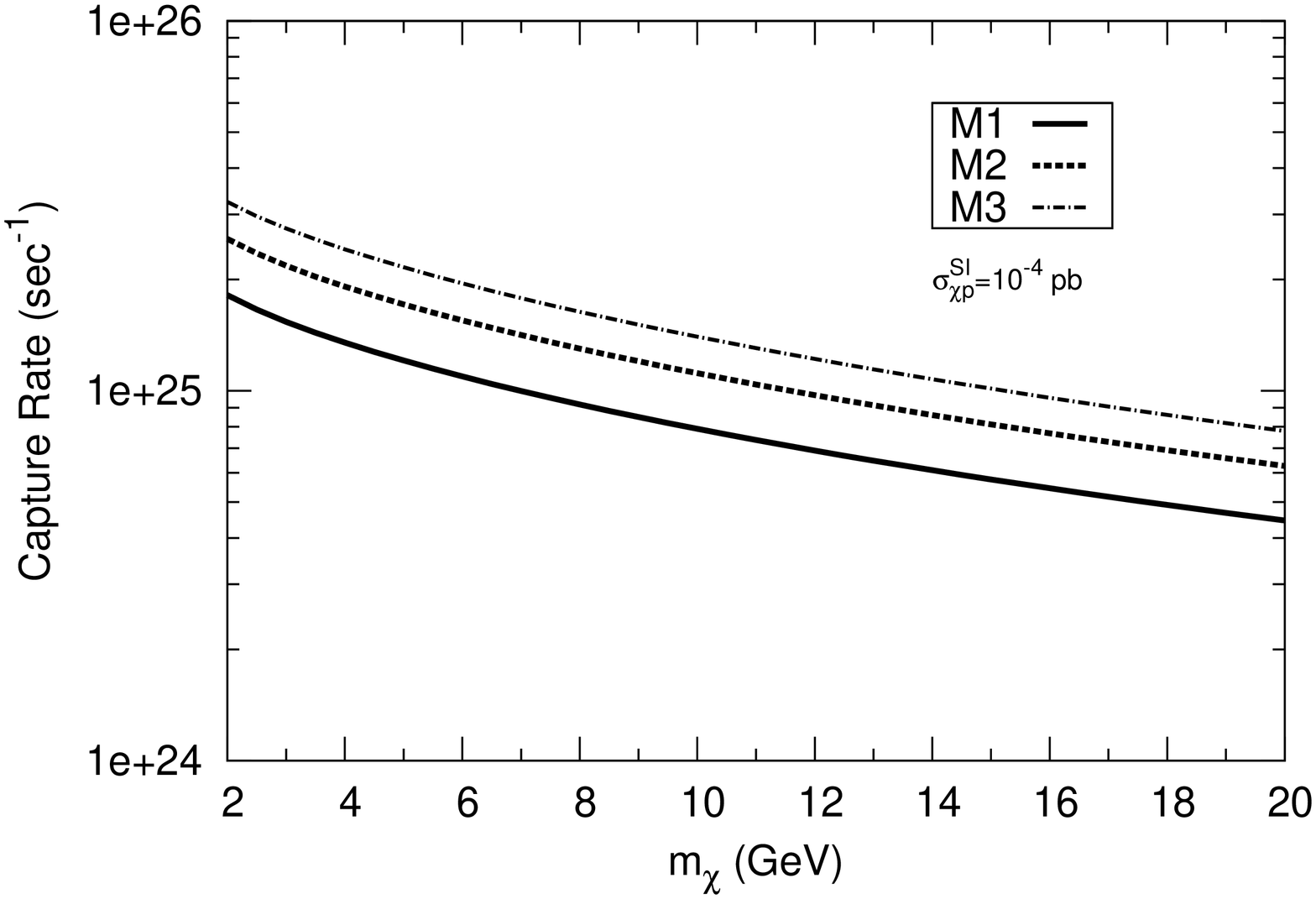,angle=270,width=3.5in}
&
\epsfig{file=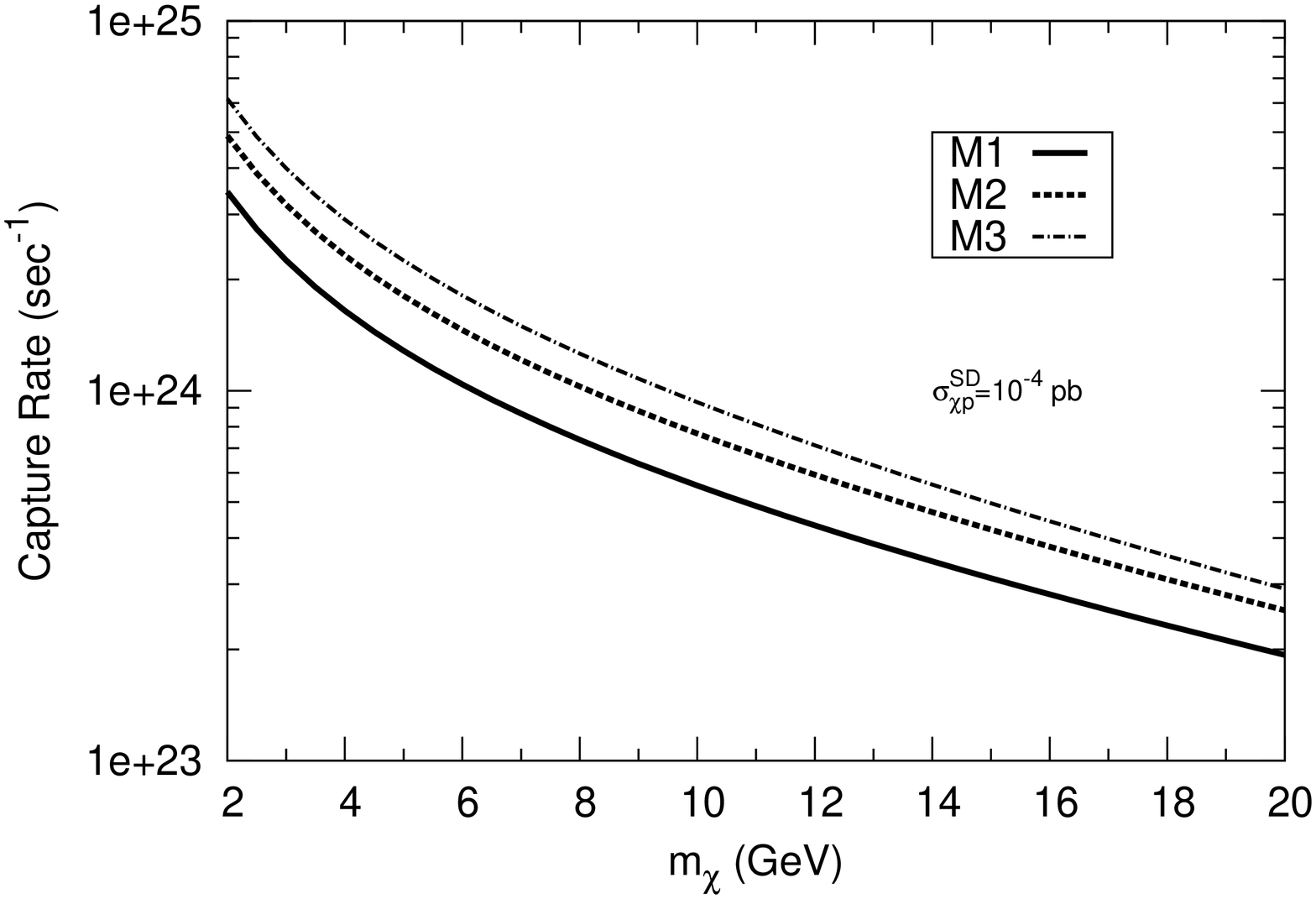,angle=270,width=3.5in}\\
\end{tabular}
\caption{The capture rate as a function of the WIMP mass for the three halo models specified 
in Table~\ref{Table:good_fit_models}, and for spin-independent (SI: left panel) and 
spin-dependent (SD: right panel) WIMP-proton interactions. All the curves are for a reference 
value of the WIMP-proton elastic SI or SD cross section of 
$10^{-4}\pb$.}
\label{Fig:capture_rates_SI_SD}
\end{figure}

\noindent
Fig.~\ref{Fig:capture_rates_SI_SD} shows the dependence of the capture rate of WIMPs by the 
Sun as a function of the WIMP's mass for the three halo models specified in 
Table~\ref{Table:good_fit_models}. As expected, for a given DM density, the capture rate 
decreases as WIMP mass increases because heavier WIMPs correspond to smaller number density of 
WIMPs. 

\begin{figure}[h]
\centering
\begin{tabular}{cc}
\epsfig{file=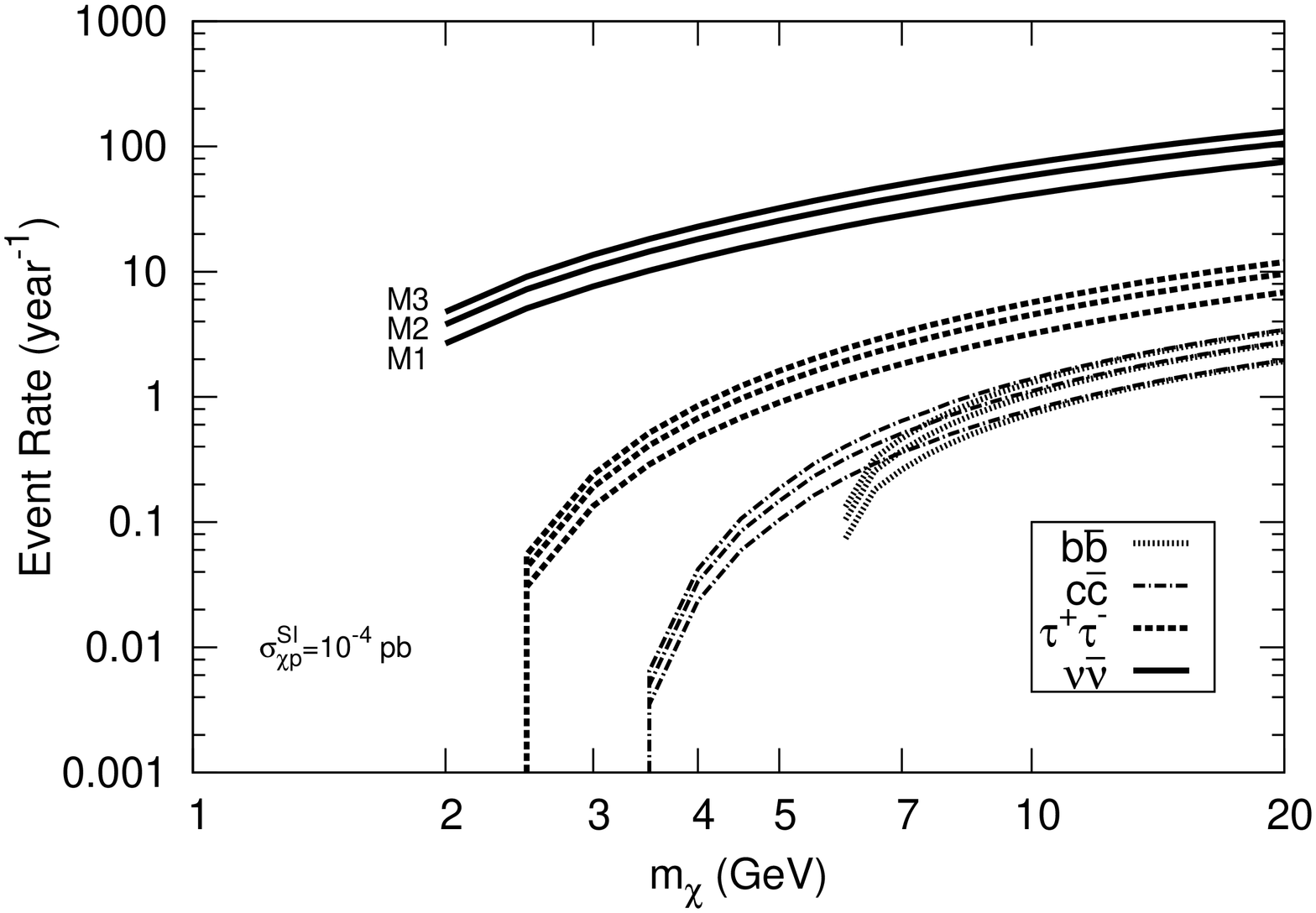,angle=270,width=3.5in}
&
\epsfig{file=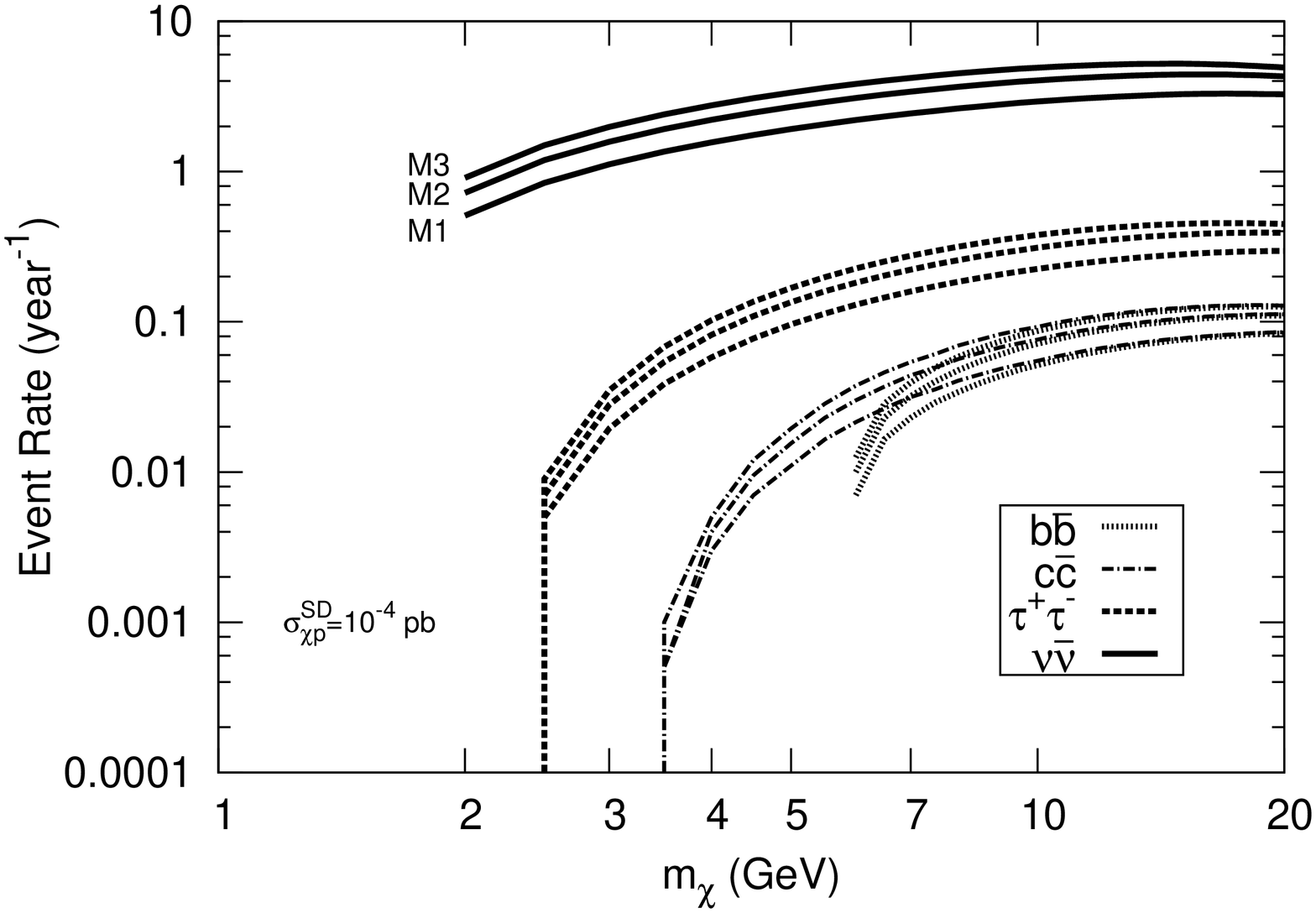,angle=270,width=3.5in}\\
\end{tabular}
\caption{ 
The upward going muon event rates in the Super-Kamiokande detector due to neutrinos from 
WIMP annihilation in the Sun as a function of the WIMP mass for the four annihilation channels 
as indicated, assuming 100\% branching ratios for each channel by itself, and for 
spin-independent (SI: left panel) and spin-dependent (SD: right panel) WIMP-proton 
interactions. The three curves for each annihilation channel correspond, as indicated, to the 
three halo models specified in Table~\ref{Table:good_fit_models}. All the curves are for a 
reference value of the WIMP-proton elastic SI or SD cross section of 
$10^{-4}\pb$.
}
\label{Fig:event_rates_SI_SD}
\end{figure}

\newpara 
The event rates in the S-K detector as a function of the WIMP mass for the four different WIMP 
annihilation channels are shown in Figure~\ref{Fig:event_rates_SI_SD} assuming 100\% 
branching ratio for each channel by itself. For each annihilation channel the three curves 
correspond, as indicated, to the three halo models specified in 
Table~\ref{Table:good_fit_models}. It is seen that the direct annihilation to the $\nu\anu$ 
channel dominates the event rate, followed by the $\tautau$ channel.  

\begin{figure}[h]
\centering
\begin{tabular}{cc}
\epsfig{file=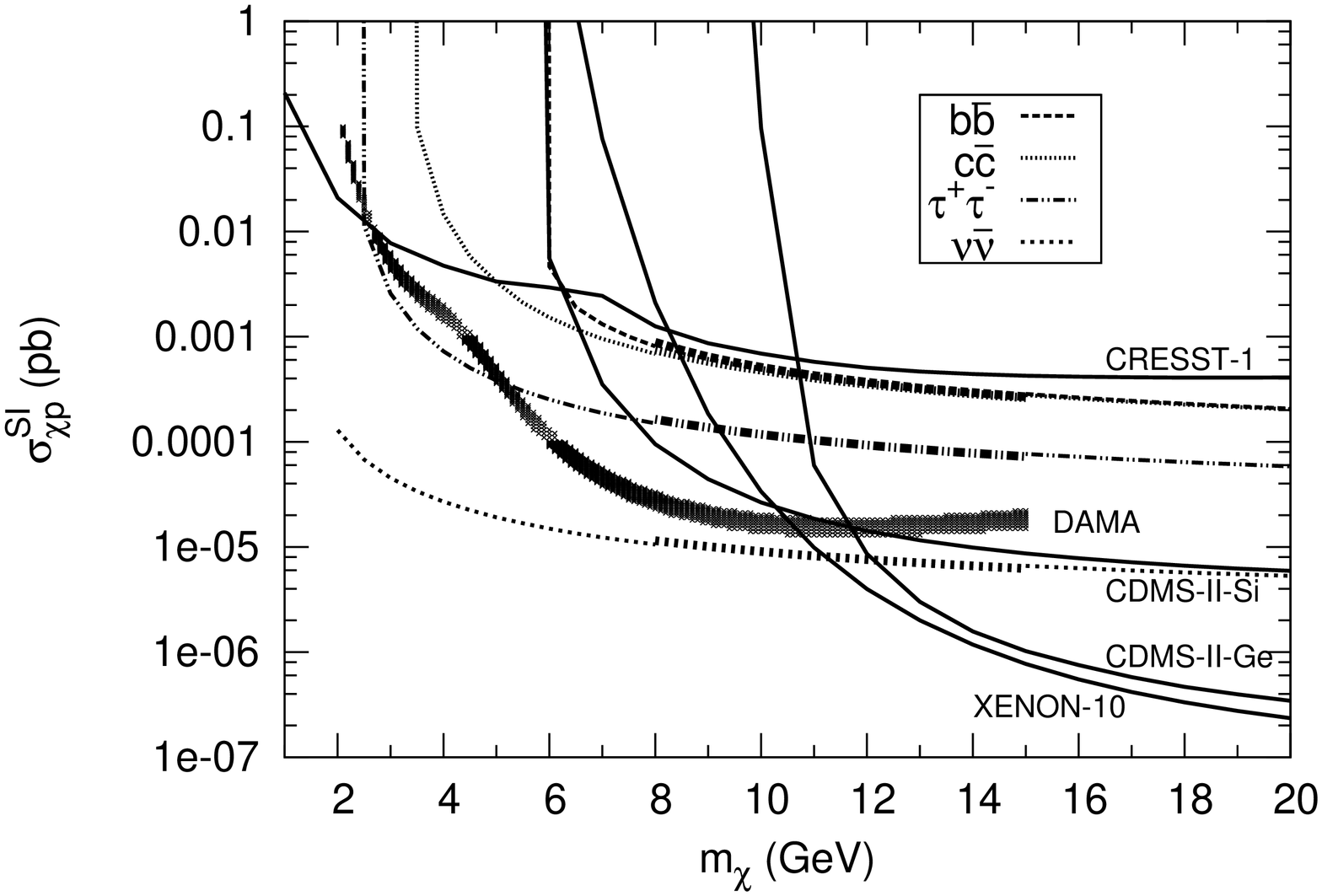,angle=270,width=3.5in}
&
\epsfig{file=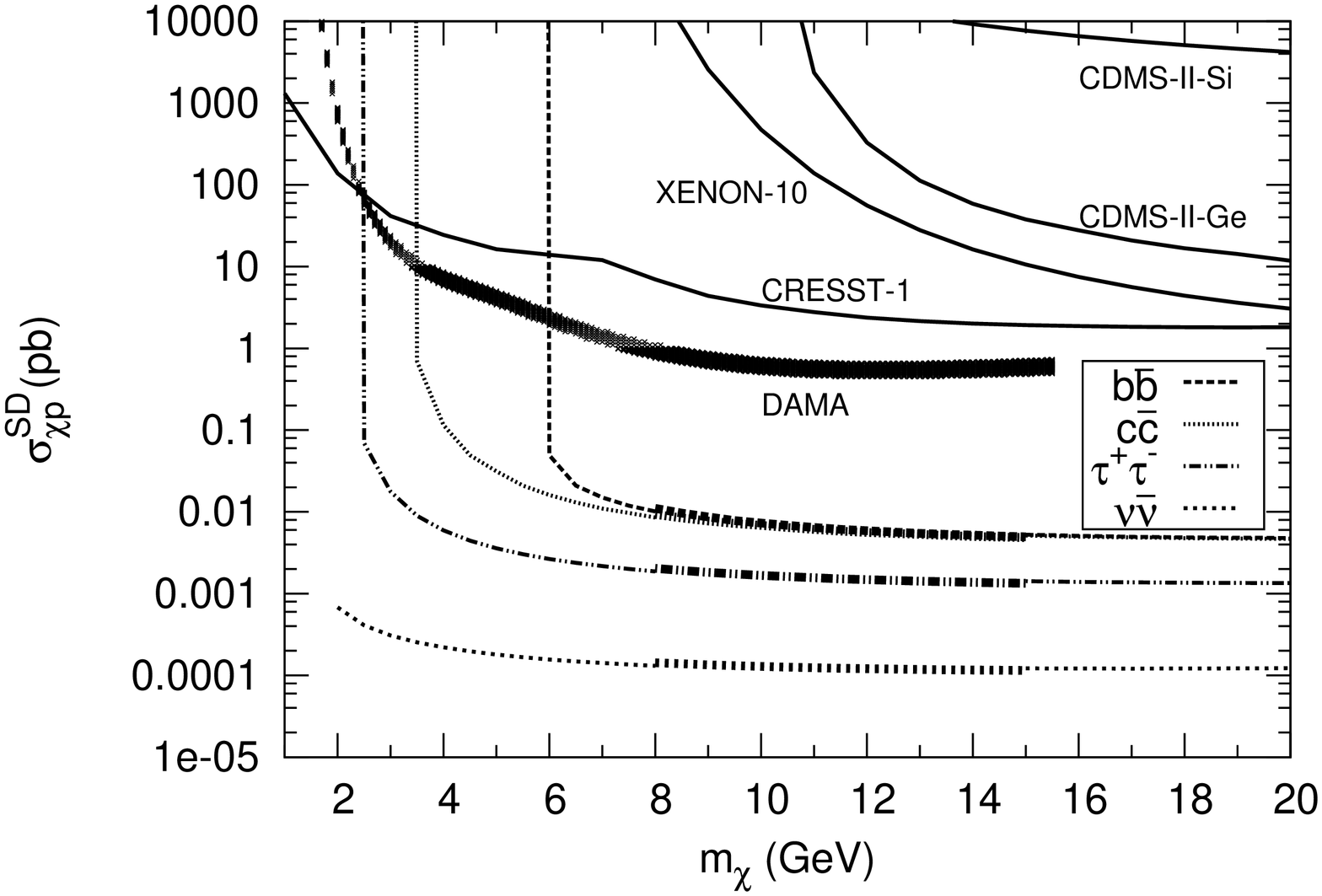,angle=270,width=3.5in}\\
\end{tabular}
\caption{The 90\% C.L.~upper limits on the WIMP-proton spin-independent (SI: left panel) and 
spin-dependent (SD: right panel) elastic cross section as a function of WIMP mass derived from
the Super-Kamiokande measurements of the up-going muon events from the direction of the 
Sun~\cite{SuperK_limit_04,Wink_1104_0679}, for the three relevant event types, namely, Fully 
Contained (FC), Stopping (S) and Through Going (TG), as discussed in the text (see 
Eq.[\ref{fc_s_tg_upper_limits}]). The thick portions of the curves serve to demarcate the 
approximate $\mchi$ ranges where the different event types make dominant contributions to the 
upper limits. The curves shown are for the four annihilation channels, assuming
100\% branching ratio for each channel by itself. The 90\% 
C.L.~allowed regions in the WIMP mass vs.~WIMP-proton elastic cross section plane 
implied by the DAMA/LIBRA experiment's claimed annual modulation signal~\cite{dama-libra} 
as well as the 90\% C.L.~upper limits on the cross section as a 
function of the WIMP mass implied by the null results from the CRESST-1~\cite{cresst-1},
CDMS-II-Si~\cite{cdmsII-Si}, CDMS-II-Ge~\cite{cdmsII-Ge} and XENON10~\cite{xenon10}
experiments (solid curves) are also shown.  All the curves shown are for our halo 
model M1 ($\rhodmsun=0.2\gev/\cm^3$) specified in Table~\ref{Table:good_fit_models}.}
\label{Fig:excl_plot_M1_SI_SD}
\end{figure}

\begin{figure}[h]
\centering
\begin{tabular}{cc}
\epsfig{file=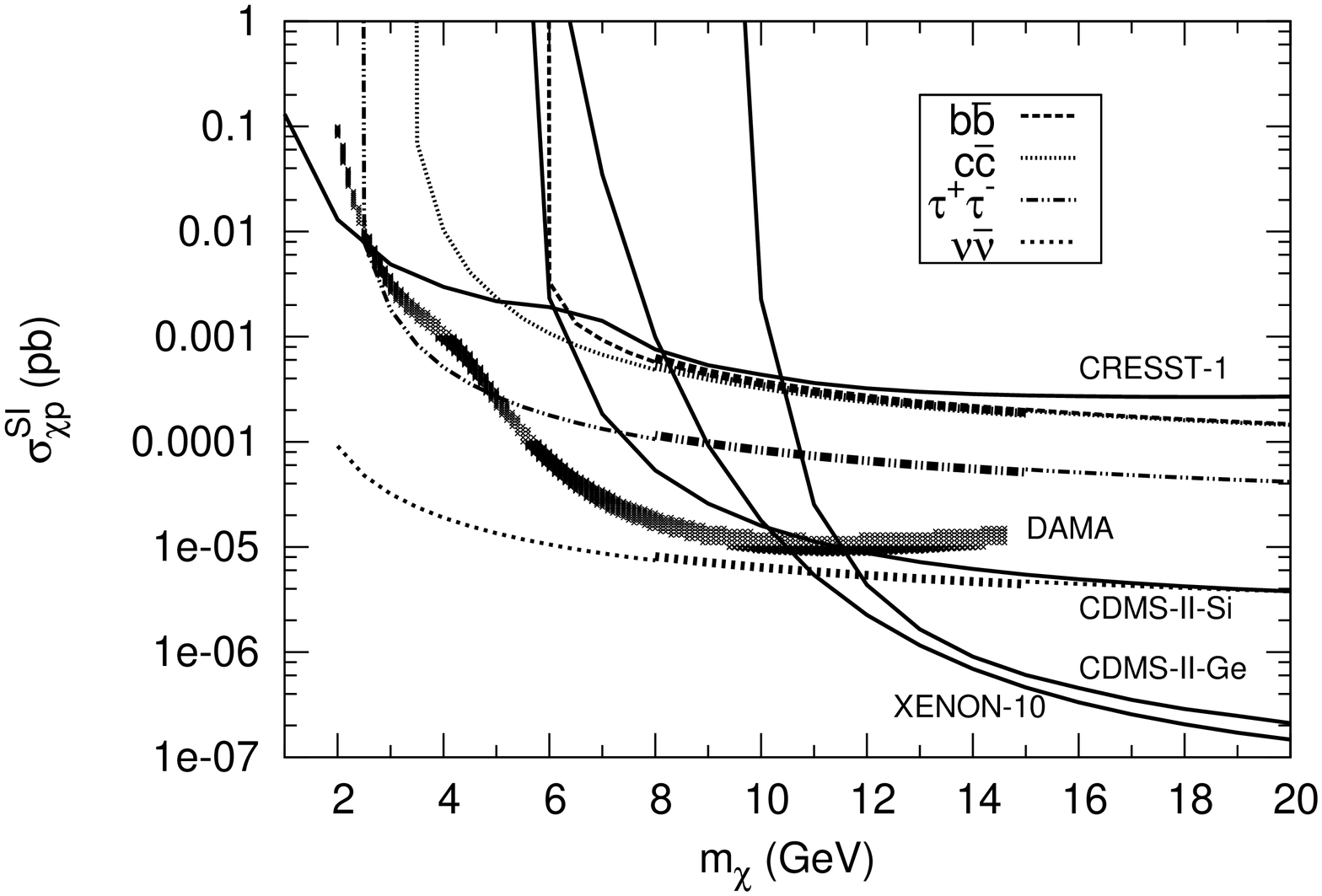,angle=270,width=3.5in}
&
\epsfig{file=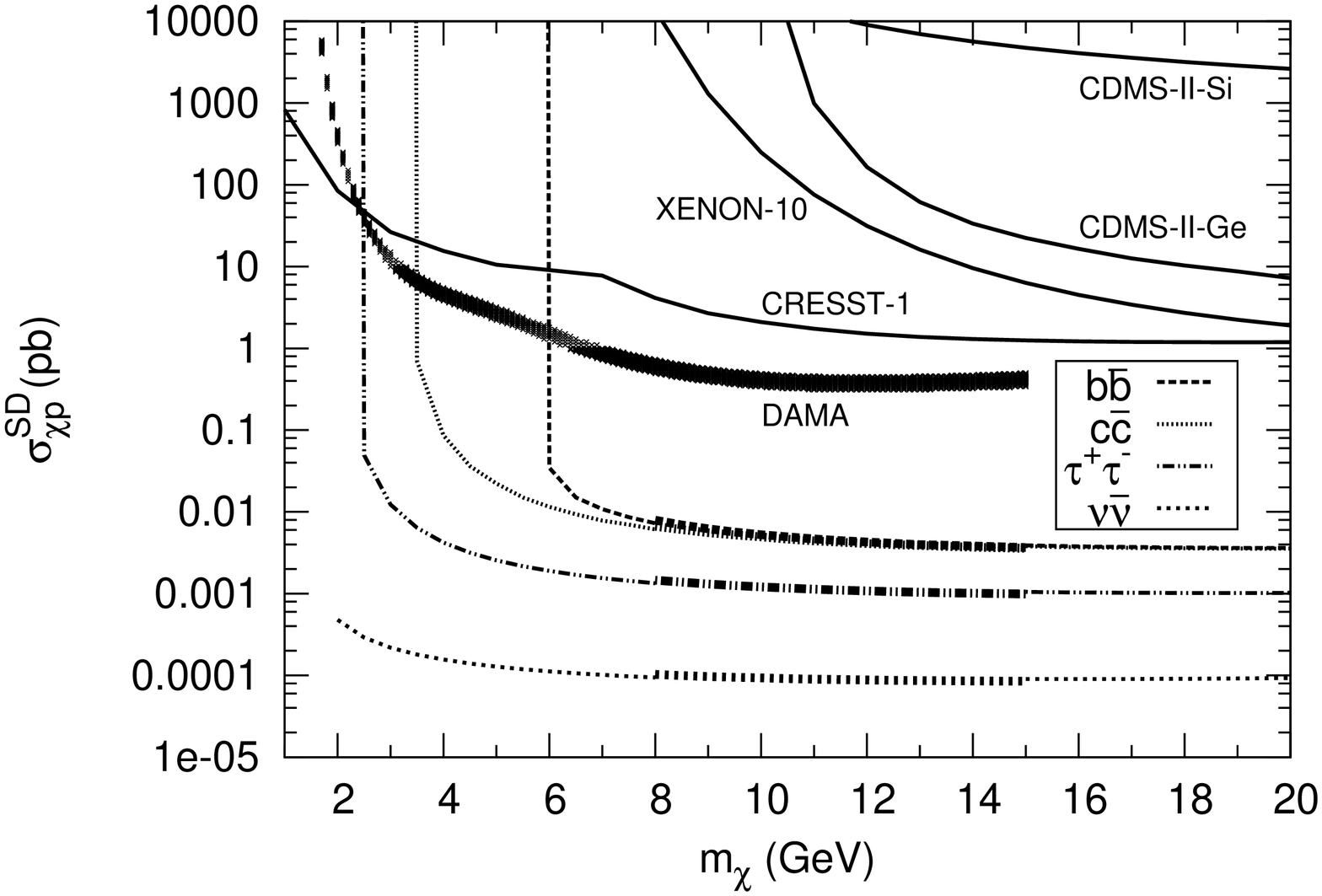,angle=270,width=3.5in}\\
\end{tabular}
\caption{Same as Fig.~\ref{Fig:excl_plot_M1_SI_SD}, but for the halo model M2
($\rhodmsun=0.3\gev/\cm^3$) specified in Table~\ref{Table:good_fit_models}.}
\label{Fig:excl_plot_M2_SI_SD}
\end{figure}

\begin{figure}[h]
\centering
\begin{tabular}{cc}
\epsfig{file=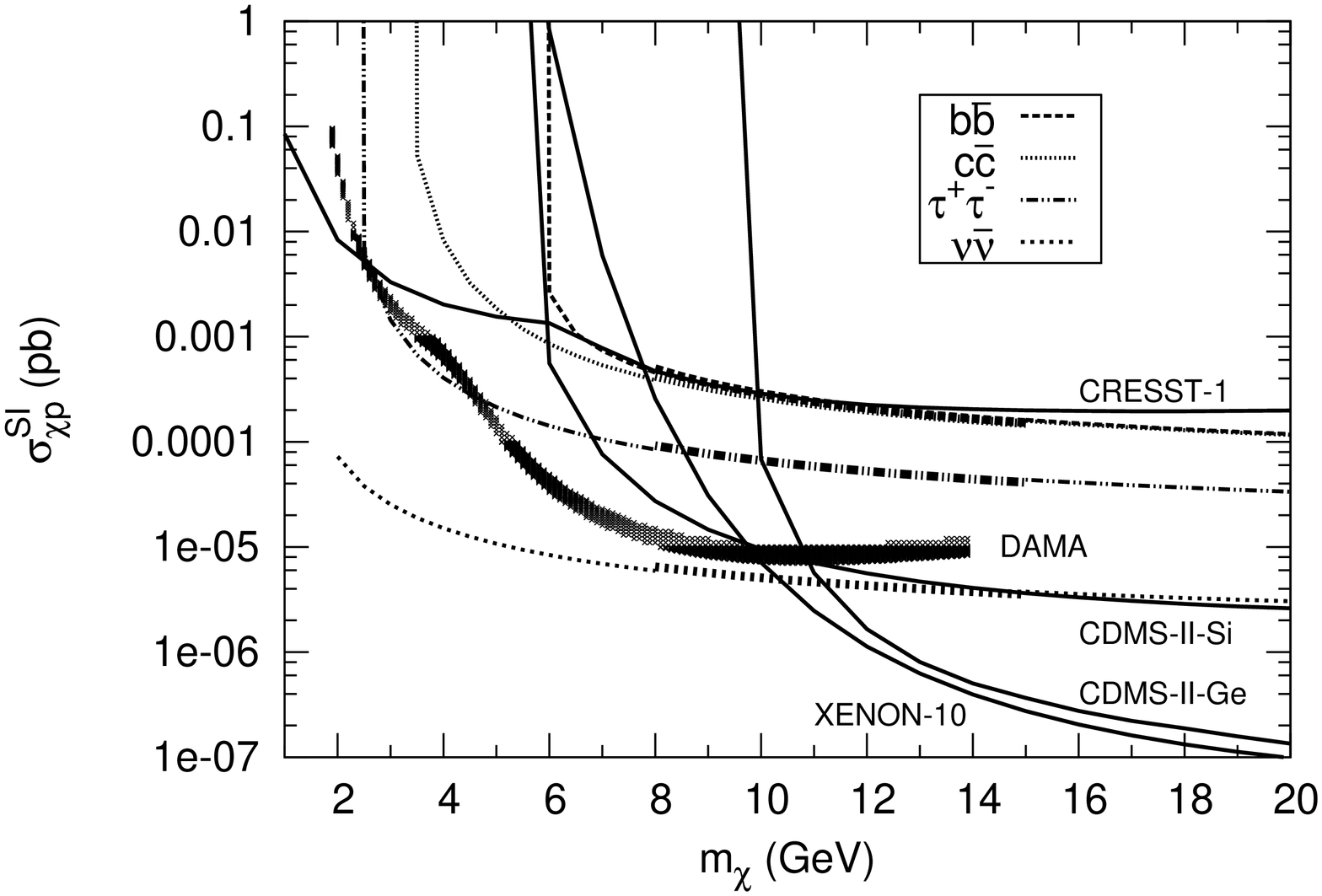,angle=270,width=3.5in}
&
\epsfig{file=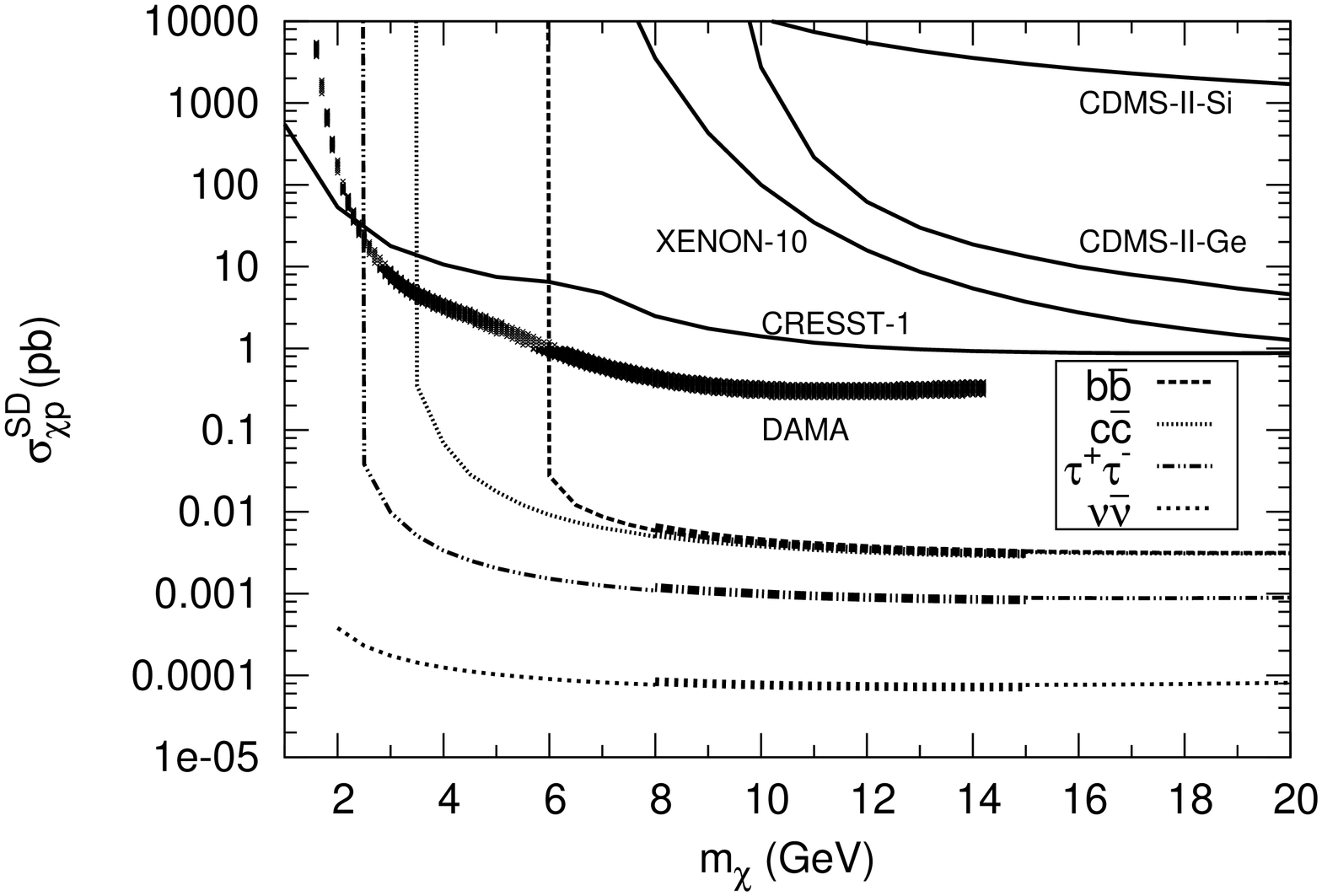,angle=270,width=3.5in}\\
\end{tabular}
\caption{Same as Fig.~\ref{Fig:excl_plot_M1_SI_SD}, but for the halo model M3
($\rhodmsun=0.4\gev/\cm^3$) specified in Table~\ref{Table:good_fit_models}.}
\label{Fig:excl_plot_M3_SI_SD}
\end{figure}

\newpara
Our main results are contained in Figures \ref{Fig:excl_plot_M1_SI_SD}, 
\ref{Fig:excl_plot_M2_SI_SD} and \ref{Fig:excl_plot_M3_SI_SD}, where we show, for the three 
halo models considered, the 90\% C.L.~upper limits 
on the WIMP-proton SI and SD elastic cross sections (as a function of WIMP mass) derived from 
the Super-Kamiokande measurements of the up-going muon events from the direction of the 
Sun~\cite{SuperK_limit_04,Wink_1104_0679}, for the four annihilation channels discussed in 
the text, assuming 
100\% branching ratio for each channel by itself. In these Figures, we also display, for the 
respective halo models, the 90\% C.L.~allowed regions~\cite{cbc_JCAP2010} in the WIMP mass 
vs.~WIMP-proton elastic cross section plane implied by 
the DAMA/LIBRA collaboration's claimed annual modulation signal~\cite{dama-libra}, as well as 
the 90\% C.L.~upper limits~\cite{cbc_JCAP2010} on the relevant cross section as a function of 
the WIMP mass implied by the null results from the CRESST-1~\cite{cresst-1}, 
CDMS-II-Si~\cite{cdmsII-Si}, CDMS-II-Ge~\cite{cdmsII-Ge} and XENON10~\cite{xenon10} 
experiments.  

\newpara
The curves in Figures \ref{Fig:excl_plot_M1_SI_SD}, \ref{Fig:excl_plot_M2_SI_SD} and 
\ref{Fig:excl_plot_M3_SI_SD} allow us to derive upper limits on the branching 
fractions of the various WIMP annihilation channels, from the requirement of consistency of 
the S-K-implied upper limits on the WIMP-proton elastic cross section with the 
DAMA-compatible regions. These upper limits are shown in 
Table \ref{Table:branching_ratios_up_lims} for the three halo models discussed in the text. 
\begin{table}[h]
\centering
\begin{tabular}{|c||c||c|c|c|c|}
\hline
\multicolumn{1}{|c||}{ } & \multicolumn{1}{|c||}{EVENT TYPE} &\multicolumn{4}{|c|}{UPPER LIMITS ON THE BRANCHING FRACTIONS (in \%)~(M1, M2, M3)}  \\ \cline{3-6}
\multicolumn{1}{|c||}{ } & \multicolumn{1}{|c||}{($\mchi$ range in \gev)} &  $\bbarb$ & $\cbarc$ & $\tautau$ & $\nu\anu$ \\
\hline
\hline
\multirow{3}{*}{~SI~} & FC $(2.0-8.0)$&$~100,100,100~$& $~100,100,100~$&$~35,40,45~$& $~0.4,0.6,0.8~$  \\ \cline{2-6}
&~FC+S $(8.0-15.0)$~  &$~100,100,100~$&$~100,100,100~$&$~100,100,100~$&$~25,30,35~$  \\ \cline{2-6}
& ~FC+S+TG $(15.0-20.0)$~  &$~100,100,100~$&$~100,100,100~$&$~100,100,100~$&$~25,30,35~$  \\ \hline \hline
\multirow{3}{*}{~SD~} & ~FC ~$(2.0-8.0)$ & $~0.5,0.6,0.7~$  & $~0.5,0.6,0.7~$  & $~0.05,0.06,0.07~$  & $~0.0005,0.0006,0.0007~$  \\ \cline{2-6}
& ~FC+S $(8.0-15.0)$~  & $~0.6,0.7,0.8~$  & $~0.6,0.7,0.8~$  & $~0.14,0.16,0.18~$  & $~0.012,0.014,0.016~$   \\ \cline{2-6}
& ~FC+S+TG $(15.0-20.0)$~   & $~0.6,0.7,0.8~$  & $~0.6,0.7,0.8~$  & $~0.14,0.16,0.18~$  & $~0.012,0.014,0.016~$  \\ \cline{2-6}
\hline
\end{tabular}
\caption{Upper limits --- derived from Figures \ref{Fig:excl_plot_M1_SI_SD}, 
\ref{Fig:excl_plot_M2_SI_SD} and \ref{Fig:excl_plot_M3_SI_SD} --- on the branching fractions 
for the four annihilation channels, from the requirement of consistency of the S-K implied 
upper limits on the WIMP-proton elastic cross sections with the ``DAMA-compatible" region of 
the WIMP mass versus cross section parameter space (within which the annual 
modulation signal observed by the DAMA/LIBRA experiment~\cite{dama-libra} is compatible 
with the null results of other DD experiments determined within the context of our halo 
model~\cite{cbc_JCAP2010}), for both spin-independent (SI) and spin-dependent (SD) 
interactions and the three halo models specified in Table~\ref{Table:good_fit_models}. The 
limits are calculated using the three different upward-going muon event types, namely, Fully 
Contained (FC), Stopping (S) and Through Going (TG). The three consecutive numbers for each 
annihilation channel and muon event type refer to the three different halo models M1, M2, M3,   
as indicated.}
\label{Table:branching_ratios_up_lims}
\end{table}
\newpara
Clearly, for the case of spin-independent interaction, there are no constraints on the 
branching fractions for the $\bbarb$ and $\cbarc$ channels since the DAMA-compatible region is 
already consistent 
with the S-K upper limit even for 100\% branching fractions in these channels (the respective 
curves for the various annihilation channels only move upwards, keeping the shape same, as the 
branching fractions are made smaller). At the same time, for the $\tautau$ channel and SI 
interaction, although a 100\% branching fraction in this channel allows a part of the 
DAMA-compatible region to be consistent with the S-K upper 
limit, consistency of the {\it entire} DAMA-compatible region with the S-K upper 
limit requires the branching fraction for this channel to be less than 35--45\% depending on 
the halo model. On the other hand, for the $\nu\anu$ channel and SI interaction, there are 
already strong upper limits (at the level of 25 -- 35\%) on the branching fraction for this 
channel for consistency of even a part of the DAMA-compatible region with the S-K upper limit; 
and these upper limits become significantly more stringent (by about two orders of magnitude) 
if the entire DAMA-compatible region is required to be consistent with the S-K upper limits. 

\newpara 
The constraints on the branching fractions of various annihilation channels are, however,  
much more severe in the case of spin-dependent interaction: For the quark 
channels, only parts of the DAMA-compatible region can be made consistent with the S-K upper 
limits, and that only if the branching fractions for these channels are restricted at the 
level of (0.6 -- 0.8)\%. On the other hand, for $\tautau$ and $\nu\anu$ channels, 
parts of the 
DAMA-compatible regions can be consistent with S-K upper limits only if their branching 
fractions are restricted at the level of (0.14 -- 0.18)\% and (0.012 -- 0.016)\%, 
respectively, while consistency of the entire DAMA-compatible regions 
with the S-K upper limits requires these fractions to be respectively lower by about a factor 
of 2.5 (for the $\tautau$ channel) and a factor of about 25 (for the $\nu\anu$ channel). 

\newpara
The above small numbers for the upper limits on the branching fractions of the four dominant 
neutrino producing WIMP annihilation channels imply, in the case of spin-dependent WIMP 
interaction, that the DAMA-allowed region of the $\mchi$ -- $\sigmachipSD$ parameter space is 
essentially ruled out by the S-K upper limit on neutrinos from possible WIMP annihilations in 
the Sun, unless, of course, WIMPs efficiently evaporate from the Sun --- which may be the case 
for relatively small mass WIMPs below 4 GeV~\cite{Hooper_Petriello_Zurek_Kamion_PRD_09} --- or 
there are other non-standard but dominant WIMP annihilation channels that somehow do not 
eventually produce any significant number of neutrinos while restricting annihilation to quark 
($\bbarb$, \ $\cbarc\,$) channels to below 0.5\% level and $\tautau$ and 
$\nu\anu$ channels to below 0.05\% and 0.0005\% level, respectively. In the case of 
spin-independent interaction, however, the DAMA-compatible region of the $\mchi$ -- 
$\sigmachipSI$ parameter space (or at least a part thereof) remains unaffected by the S-K 
upper limit if WIMPs annihilate dominantly to quarks and/or tau leptons, and annihilation 
directly to neutrinos is restricted below $\sim$ (25 -- 35)\% level. At the same time, 
portions of the 
DAMA-compatible region can be excluded if WIMP annihilation to $\tautau$ occurs at larger than 
(35 -- 45)\% level and/or annihilation to $\nu\anu$ occurs at larger than (0.4 -- 0.8)\% 
level. These results, based as they are on a self-consistent model of the Galaxy's dark 
matter halo, 
the parameters of which are determined by a fit to the rotation curve of the Galaxy, 
strengthen, at the qualitative level, the earlier conclusion within the SHM  
that the S-K upper limit on the possible flux of neutrinos due to WIMP 
annihilation in the Sun severely restricts the DAMA region of the WIMP mass versus cross 
section plane, especially in the case of spin-dependent interaction of WIMPs with nuclei, 
although the quantitative restrictions on the WIMP cross section and branching fractions of 
various WIMP annihilation channels obtained here are different (in some cases by more 
than a factor of 10) from those obtained in earlier calculations within the SHM.   

\newpara
\underline{\it Note added}: After the completion of the main calculations of the present work, new 
results of the S-K collaboration's search for upward-going muons (``upmus") due to neutrinos 
from Sun~\cite{S-K_arxiv:1108.3384} have appeared. These new results are based on a data 
set consisting of 3109.6 days of data, nearly double the size of the old data set of 1679.6 
days used in Ref.~\cite{SuperK_limit_04} and in the analysis of this paper so far. 
Here we consider these new results of Ref.~\cite{S-K_arxiv:1108.3384} and the resulting 
changes to the 
constraints on various WIMP annihilation channels derived above using the earlier S-K 
results. In general, we find that with the new S-K results the upper limits on the branching 
fractions of various annihilation channels become more stringent by a factor of 3 -- 4 than 
those derived above.  

\newpara
The upmu event categories used in the new S-K paper~\cite{S-K_arxiv:1108.3384} are somewhat  
different from those in their earlier work~\cite{SuperK_limit_04}. These are: ``stopping" 
(S), ``non-showering through-going" (NSTG), and ``showering through-going" (STG); see 
Ref.~\cite{S-K_arxiv:1108.3384} for details. For a given WIMP 
mass, Figure 2 of Ref.~\cite{S-K_arxiv:1108.3384} allows us to read out the fraction of each 
upmu event type contributing to the total number of events, from the 
consideration that the typical maximum energy of a neutrino produced in the annihilation of 
a WIMP of mass $\mchi$ is $\sim\mchi/2$. For low WIMP masses of our interest in this paper, 
$\mchi\lsim 20\gev$ (and hence typical neutrino energies $\lsim 10\gev$), the stopping 
events dominate and constitute more than 70\% of the total number of upmu events, as clear 
from Figure 2 of Ref.~\cite{S-K_arxiv:1108.3384}. It is thus expected, as indeed we do find 
from our calculations, that the most stringent upper limits on the branching fractions of 
various WIMP annihilation channels for low WIMP masses come from the observed rate of 
these Stopping events.\footnote{Recall that, for the older data set~\cite{SuperK_limit_04}, 
the most stringent upper limits came from the fully-contained (FC) events; see 
Table~\ref{Table:branching_ratios_up_lims} above.} 

\begin{figure}[h]
\centering
\begin{tabular}{cc}
\epsfig{file=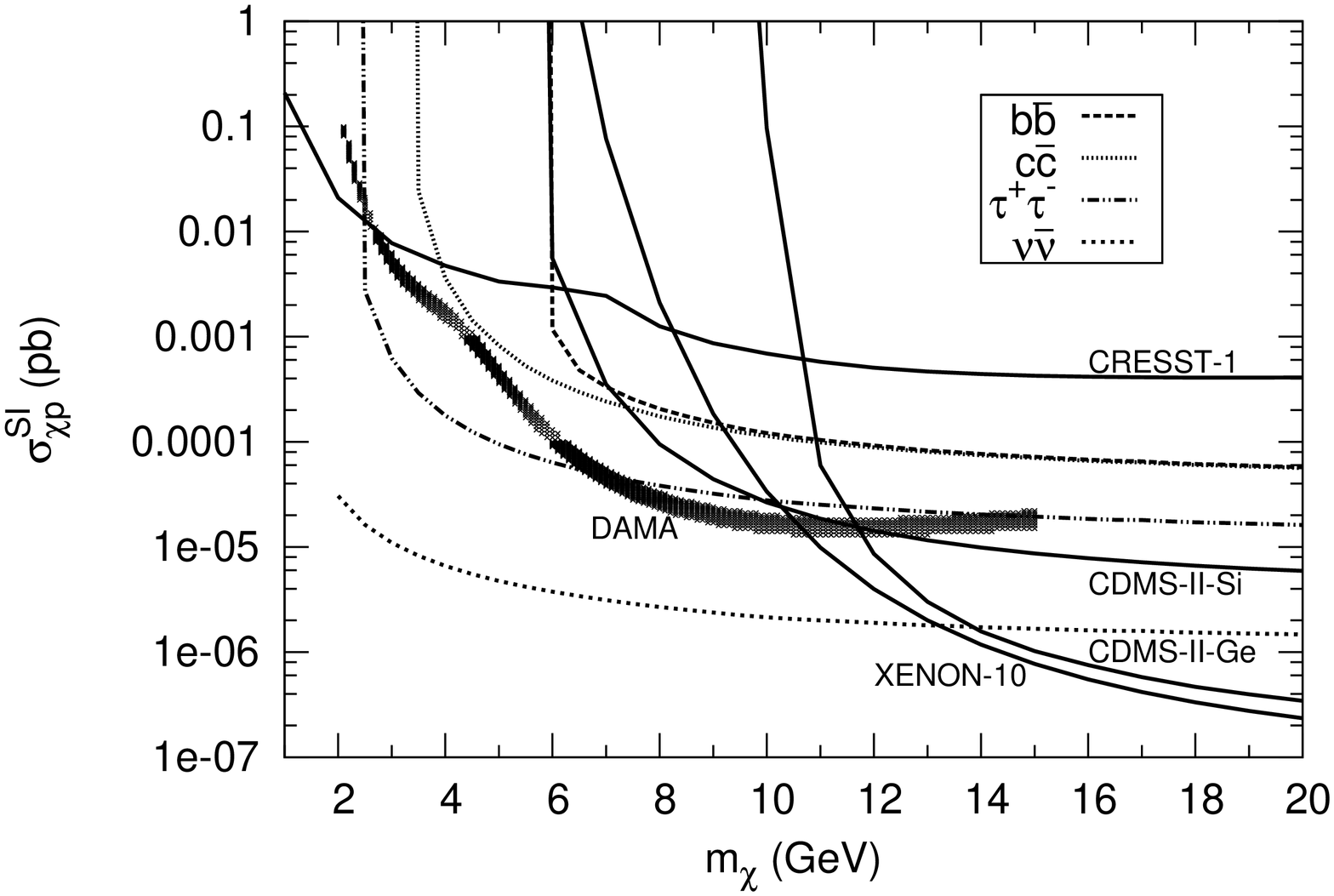,angle=270,width=3.5in}
&
\epsfig{file=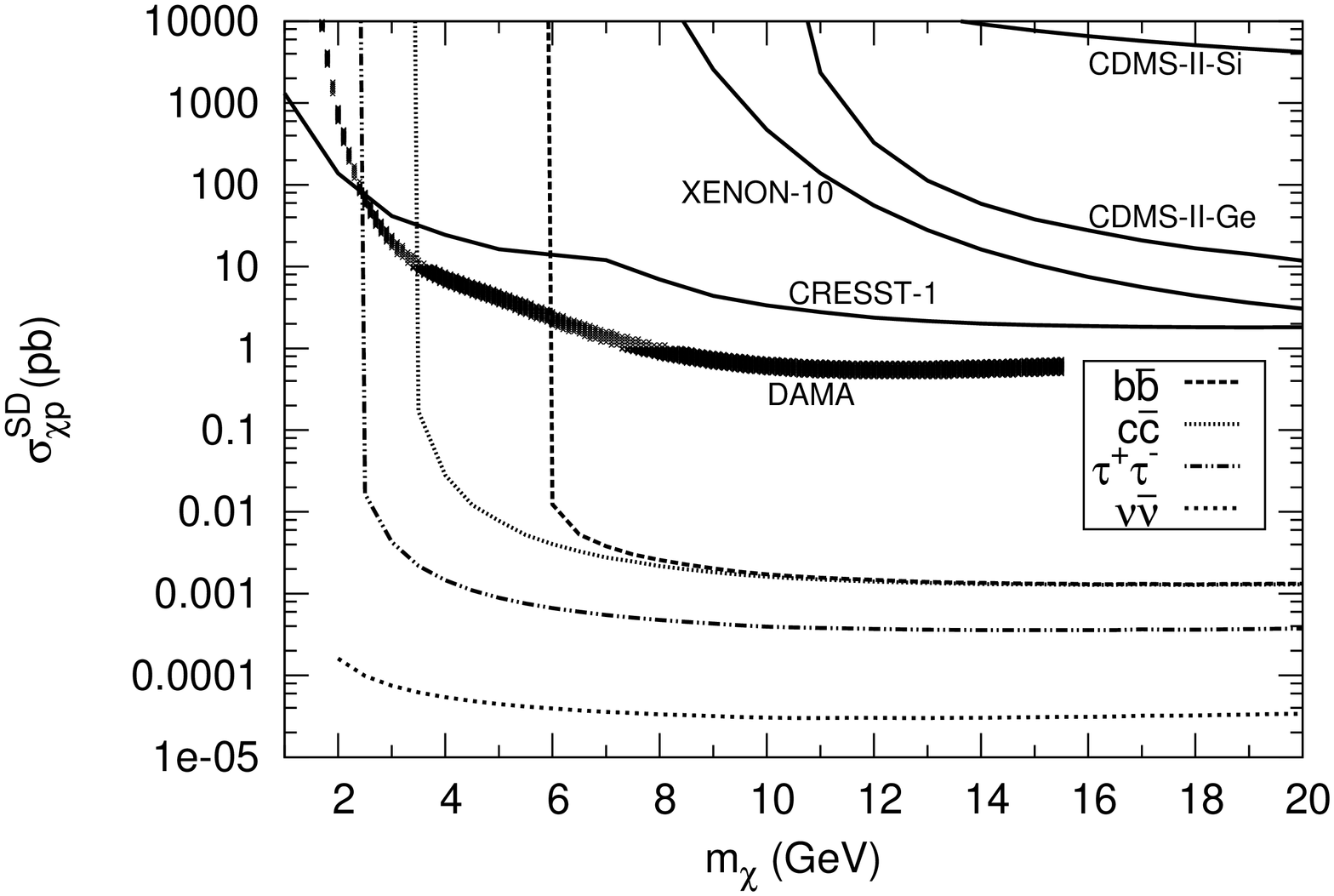,angle=270,width=3.5in}\\
\end{tabular}
\caption{Same as Fig.~\ref{Fig:excl_plot_M1_SI_SD}, but using the 
new S-K data from Ref.~\cite{S-K_arxiv:1108.3384} and considering their ``Stopping" upmu 
events only.}
\label{Fig:excl_plot_S-K_new_Tanaka_M1_SI_SD}
\end{figure}

\newpara
The 90\% C.L.~Poissonian upper limit on the rate of these Stopping-type upmu events for the 
new data set of Ref.~\cite{S-K_arxiv:1108.3384}, estimated from the total number of this 
type of upmu events and the number of background upmus due to atmospheric neutrinos given in 
Figure 3 of that reference, is~\footnote{We take the events in the 0--30 degree cone 
half-angle bin around the Sun to be consistent with the analysis done above for the earlier 
S-K data set.} $\sim 3.27 \yr^{-1}$. With this, we can calculate, as we did in the analysis 
above, the 90\% C.L.~upper limits on the WIMP-proton SI and SD elastic cross 
sections as a function of WIMP mass for the new S-K data set of 
Ref.~\cite{S-K_arxiv:1108.3384}, for the case of 100\% branching ratio for each of the four 
annihilation channels considered above. The results, for our best-fit halo model M1, are 
shown in Figure \ref{Fig:excl_plot_S-K_new_Tanaka_M1_SI_SD}.
\nopagebreak

\newpara
The resulting upper limits on the branching fractions of the four annihilation channels, 
derived from the requirement of consistency of the new S-K-implied upper limits on the 
WIMP-proton elastic cross sections shown in Figure 
\ref{Fig:excl_plot_S-K_new_Tanaka_M1_SI_SD} with the ``DAMA-compatible" regions, are 
displayed in Table \ref{Table:branching_ratios_up_lims_SKnew_Tanaka_M1}. 
\begin{table}[h]
\centering
\begin{tabular}{|c||c|c|c|c|}
\hline
\multicolumn{1}{|c|}{ } & \multicolumn{4}{|c|}{Upper limits on the branching fractions (in
\%)}  \\
\multicolumn{1}{|c|}{ } & \multicolumn{4}{|c|}{from ``Stopping" events, with halo model M1}
\\ \cline{2-5}
\multicolumn{1}{|c|}{ } &  \qquad $\bbarb$ \qquad  \qquad  &  \qquad $\cbarc$ \qquad  \qquad  &
\qquad $\tautau$ \qquad   \qquad &  \qquad $\nu\anu$  \qquad \qquad  \\
\hline
\hline
{~SI~} &  \qquad $ 100  $  \qquad  \qquad &  \qquad $   100  $  \qquad \qquad  &  \qquad $
10   $ \qquad  \qquad  &  \qquad $   0.11   $ \qquad  \qquad   \\ \cline{2-5}
\hline
{~SD~} &  \qquad $  0.12$  \qquad  \qquad  & \qquad  $ 0.12 $   \qquad \qquad  & \qquad  $
0.012 $ \qquad  \qquad   &  \qquad  $ 0.00013 $  \qquad  \qquad  \\ \cline{2-5}
\hline
\end{tabular}
\caption{Upper limits on the branching fractions of the four annihilation channels,
derived from the requirement of consistency of the new S-K-implied upper limits on the
WIMP-proton elastic cross sections shown in Figure
\ref{Fig:excl_plot_S-K_new_Tanaka_M1_SI_SD} with the ``DAMA-compatible" regions, for both 
spin-independent (SI) and spin-dependent (SD) interactions}
\label{Table:branching_ratios_up_lims_SKnew_Tanaka_M1}
\end{table}
A comparison with the corresponding numbers given in Table 
\ref{Table:branching_ratios_up_lims} shows that the new upper limits on the branching 
fractions for the relevant annihilation channels are roughly a factor of 3--4 more 
stringent.  

\section{Summary}\label{sec:Summary}
\noindent
Several studies in recent years have brought into focus the possibility that the dark 
matter may be in the form of a relatively light WIMP of mass in the few GeV range. Such 
light WIMPs with suitably chosen values of the WIMP-nucleon SI or SD elastic cross section can 
be consistent with the annual modulation signal seen in the DAMA/LIBRA 
experiment~\cite{dama-libra} without conflicting with the null results of other 
direct-detection experiments. To further probe the ``DAMA-compatible" regions of the WIMP 
parameter space --- the regions of the WIMP mass versus cross section parameter space 
within which the annual modulation signal observed by the DAMA/LIBRA experiment is compatible 
with the null results of other DD experiments --- we have studied in this paper the 
independent constraints on the WIMP-proton SI as well as SD elastic scattering cross section 
imposed by the upper limit on the neutrino flux from WIMP annihilation in the Sun given by the 
Super-Kamiokande experiment~\cite{SuperK_limit_04,Wink_1104_0679}. Assuming approximate 
equilibrium between 
the capture and annihilation rates of WIMPs in the Sun, we have calculated the 90\% 
C.L.~upper limits on the WIMP-proton SI and SD elastic cross sections as a function of the 
WIMP mass for various WIMP annihilation channels using the Super-Kamiokande upper limits, and 
examined the consistency of those limits with the 90\% C.L.``DAMA-compatible" regions. This we 
have done within the context of a self-consistent phase-space model of the finite-size dark 
matter halo of the Galaxy, namely, the Truncated Isothermal Model 
(TIM)~\cite{crbm_NewAstron_2007,cbc_JCAP2010}, in which we take into account  
the mutual gravitational interaction between the dark matter and the observed visible matter 
in a self-consistent manner, with the parameters of the model determined by a fit to the 
observed rotation curve data of the Galaxy. 

\newpara
We find that the requirement of consistency of the S-K~\cite{SuperK_limit_04,Wink_1104_0679} 
implied upper limits on the 
WIMP-proton elastic cross section as a function of WIMP mass imposes stringent restrictions 
on the branching fractions of the various WIMP 
annihilation channels. In the case of spin-independent WIMP-proton 
interaction, the S-K upper limits do not place additional restrictions on the 
DAMA-compatible region of the WIMP parameter space if the WIMPs annihilate dominantly to 
$\bbarb$ and  $\cbarc$, and if direct annihilations
to $\tautau$ and neutrinos are restricted to below $\sim$ (35 -- 45)\% and (0.4 -- 
0.8)\%, respectively. In the case of spin-dependent interactions, on the other hand, the 
restrictions on the branching fractions of various annihilation channels are much more 
stringent, essentially ruling out the DAMA-compatible region of the WIMP parameter space if 
the relatively low-mass WIMPs under consideration annihilate predominantly to any mixture of  
$\bbarb$, \ $\cbarc$, \ $\tautau$, \ and $\nu\anu$ final states. The very latest results from the 
S-K Collaboration~\cite{S-K_arxiv:1108.3384} put the above conclusions on an even firmer 
footing by making the above constraints on the branching fractions of various WIMP 
annihilation channels more stringent by roughly a factor of 3--4. Similar conclusions were 
reached earlier~\cite{Hooper_Petriello_Zurek_Kamion_PRD_09,Wink_1104_0679}
within the context of the SHM. The quantitative restrictions on the branching fractions for 
various WIMP annihilation channels obtained here and as given in 
Table~\ref{Table:branching_ratios_up_lims} (and in Table 
\ref{Table:branching_ratios_up_lims_SKnew_Tanaka_M1} for the latest S-K 
results~\cite{S-K_arxiv:1108.3384}) are, however, significantly different from those 
in the earlier works. 

\newpara
An important aspect of the Truncated Isothermal model of the Galactic halo used in the 
present calculation is the {\it 
non-Maxwellian nature of the WIMP velocity distribution} in this model, as opposed to the 
Maxwellian distribution in the SHM (see Ref.~[15] for details). This directly affects the 
WIMP capture rate (and consequently the annihilation rate), resulting in 
significant quantitative differences in the values of the upper limits on 
the WIMP-proton elastic cross sections (implied by the S-K upper limits on the neutrinos 
from the Sun) compared to the values in the SHM. Similarly, the upper limits on the 
branching fractions of various possible WIMP annihilation channels (from the requirement of 
compatibility with DAMA results) are also changed. At a qualitative level, however, 
the general conclusion reached 
earlier~\cite{Hooper_Petriello_Zurek_Kamion_PRD_09,Wink_1104_0679}
within the context of the SHM --- that S-K upper limits on neutrinos from the Sun 
severely restrict the DAMA-compatible region of the WIMP parameter space --- remains
true in the present model too, thus adding robustness to this conclusion.  

%
\section*{Acknowledgments}
We thank Soumini Chaudhury, Martin Winkler and Dan Hooper for useful discussions and 
communications. 

\end{document}